\documentclass[twocolumn]{aastex63}
\newcommand{\caii}{\hbox{Ca\,{\scriptsize II}}}

\newcommand{\feh}{\hbox{[Fe/H]}}

\newcommand{\hbeta}{\hbox{H$\beta$}}
\newcommand{\hgamma}{\hbox{H$\gamma$}}
\newcommand{\hdelta}{\hbox{H$\delta$}}
\newcommand{\caiik}{\hbox{Ca\,{\scriptsize II}\,K}}

\newcommand{\uu}{\hbox{\it U\/}}
\newcommand{\bb}{\hbox{\it B\/}}
\newcommand{\vv}{\hbox{\it V\/}}
\newcommand{\rr}{\hbox{\it R\/}}
\newcommand{\ii}{\hbox{\it I\/}}
\newcommand{\G}{\hbox{\it G\/}}
\newcommand{\BP}{\hbox{\it G$_{\rm BP}$\/}}
\newcommand{\RP}{\hbox{\it G$_{\rm RP}$\/}}

\newcommand{\jj}{\hbox{\it J\/}}
\newcommand{\hh}{\hbox{\it H\/}}
\newcommand{\kk}{\hbox{\it K$_s$\/}}

\newcommand{\usdss}{\hbox{\it u\/}}
\newcommand{\gsdss}{\hbox{\it g\/}}
\newcommand{\rsdss}{\hbox{\it r\/}}
\newcommand{\isdss}{\hbox{\it i\/}}
\newcommand{\zsdss}{\hbox{\it z\/}}

\newcommand{\fuv}{\hbox{\it FUV\/}}
\newcommand{\nuv}{\hbox{\it NUV\/}}

\newcommand{\ebv}{\hbox{E(\bb--\vv)\/}}

\newcommand{\cubi}{\hbox{{\it c}$_{\rm U,B,I}$\/}}
\newcommand{\deltaS}{\hbox{$\Delta$S}}

\accepted{August 6, 2019}

\shorttitle{RR Lyrae as Galactic probes: I.}
\shortauthors{M. Fabrizio et al.}

\begin{document}
\title{On the use of field RR Lyrae as Galactic probes: I. \\
The Oosterhoff dichotomy based on fundamental variables.\footnote{Partially based on
observations collected under ESO programmes 297.D-5047 (PI. G. Bono) and 083.B-0281 (PI. D. Romano).}}

\correspondingauthor{Michele Fabrizio}
\email{michele.fabrizio@ssdc.asi.it}

\author[0000-0001-5829-111X]{M.~Fabrizio}
\affil{INAF - Osservatorio Astronomico di Roma, Via Frascati 33, 00078, Monte Porzio Catone (Roma), Italy}
\affil{Space Science Data Center - ASI, Via del Politecnico s.n.c., 00133 Roma, Italy}
\author{G.~Bono}
\affil{Department of Physics, Universit\'a di Roma Tor Vergata, via della Ricerca Scientifica 1, 00133 Roma, Italy}
\affil{INAF - Osservatorio Astronomico di Roma, Via Frascati 33, 00078, Monte Porzio Catone (Roma), Italy}

\author{V.F.~Braga}
\affil{Instituto Milenio de Astrof\'isica, Santiago, Chile}
\affil{Departamento de F\'isica, Facultad de Ciencias Exactas, Universidad Andr\'es Bello, Fern\'andez Concha 700, Las Condes, Santiago, Chile}

\author{D.~Magurno}
\affil{Department of Physics, Universit\'a di Roma Tor Vergata, via della Ricerca Scientifica 1, 00133 Roma, Italy}
\affil{INAF - Osservatorio Astronomico di Roma, Via Frascati 33, 00078, Monte Porzio Catone (Roma), Italy}

\author{S.~Marinoni}
\affil{INAF - Osservatorio Astronomico di Roma, Via Frascati 33, 00078, Monte Porzio Catone (Roma), Italy}
\affil{Space Science Data Center - ASI, Via del Politecnico s.n.c., 00133 Roma, Italy}

\author{P.M.~Marrese}
\affil{INAF - Osservatorio Astronomico di Roma, Via Frascati 33, 00078, Monte Porzio Catone (Roma), Italy}
\affil{Space Science Data Center - ASI, Via del Politecnico s.n.c., 00133 Roma, Italy}

\author{I.~Ferraro}
\affil{INAF - Osservatorio Astronomico di Roma, Via Frascati 33, 00078, Monte Porzio Catone (Roma), Italy}

\author{G.~Fiorentino}
\affil{INAF-Osservatorio di Astrofisica e Scienza dello Spazio di Bologna, Via Piero Gobetti 93/3, I-40129, Bologna, Italy}

\author{G.~Giuffrida}
\affil{INAF - Osservatorio Astronomico di Roma, Via Frascati 33, 00078, Monte Porzio Catone (Roma), Italy}

\author{G.~Iannicola}
\affil{INAF - Osservatorio Astronomico di Roma, Via Frascati 33, 00078, Monte Porzio Catone (Roma), Italy}

\author{M.~Monelli}
\affil{Instituto de Astrof\'{i}sica de Canarias, Calle Via Lactea s/n, E-38200 La Laguna, Tenerife, Spain}
\affil{Departamento de Astrof\'{i}sica, Universidad de La Laguna, E-38200 La Laguna, Tenerife, Spain}

\author{G.~Altavilla}
\affil{INAF - Osservatorio Astronomico di Roma, Via Frascati 33, 00078, Monte Porzio Catone (Roma), Italy}
\affil{Space Science Data Center - ASI, Via del Politecnico s.n.c., 00133 Roma, Italy}

\author{B.~Chaboyer}
\affil{Department of Physics and Astronomy, Dartmouth College, Hanover, NH 03755, USA}

\author{M.~Dall'Ora}
\affil{INAF-Osservatorio Astronomico di Capodimonte, Salita Moiariello 16, 80131 Napoli, Italy}

\author{C.K.~Gilligan}
\affil{Department of Physics and Astronomy, Dartmouth College, Hanover, NH 03755, USA}

\author{A.~Layden}
\affil{Bowling Green State University, Bowling Green, OH 43403, USA}

\author{M.~Marengo}
\affil{Department of Physics and Astronomy, Iowa State University, Ames, IA 50011, USA}

\author{M.~Nonino}
\affil{INAF-Osservatorio Astronomico di Trieste, Via G.B. Tiepolo, 11, I-34143 Trieste, Italy}

\author{G.W.~Preston}
\affil{Carnegie Observatories, 813 Santa Barbara Street, Pasadena, CA 91101, USA}

\author{B.~Sesar}
\affil{Deutsche B\"orse AG, Mergenthalerallee 61, 65760 Eschborn, Germany}

\author{C.~Sneden}
\affil{Department of Astronomy and McDonald Observatory, The University of Texas, Austin, TX 78712, USA}

\author{E.~Valenti}
\affil{European Southern Observatory, Karl-Schwarzschild-Str. 2, 85748 Garching bei M\"unchen, Germany}

\author{F.~Th\'evenin}
\affil{Universit\'{e} de Nice Sophia-antipolis, CNRS, Observatoire de la C\^{o}te d'Azur, Laboratoire Lagrange, BP 4229, F-06304 Nice, France}

\author{E.~Zoccali}
\affil{Instituto Milenio de Astrof\'isica, Santiago, Chile}
\affil{Pontificia Universidad Cat\'olica de Chile, Instituto de Astrof\'isica, Av. Vicu\~na Mackenna 4860, 7820436, Macul, Santiago, Chile}

%_____________________________________________________________________
\begin{abstract}
We collected a large data set of field RR Lyrae stars (RRLs) by using catalogues
already available in the literature and {\it Gaia} DR2.
We estimated the iron abundances for a sub-sample of 2,382 fundamental RRLs 
(\deltaS\ method: \caiik, \hbeta, \hgamma\ and \hdelta\ lines) for which are 
publicly available medium-resolution SDSS-SEGUE spectra. 
We also included similar estimates available in the literature
ending up with the largest and most homogeneous spectroscopic data set ever
collected for RRLs (2,903). The metallicity scale was validated by using iron
abundances based on high resolution spectra for a fundamental field RRL (V~Ind),
for which we collected X-shooter spectra covering the entire pulsation cycle. The
peak (\feh=--1.59$\pm$0.01) and the standard deviation ($\sigma$=0.43~dex) of the metallicity
distribution agree quite well with similar estimates available in the
literature. The current measurements disclose a well defined metal-rich tail
approaching Solar iron abundance.    
The spectroscopic sample plotted in the Bailey diagram (period \textit{vs} luminosity
amplitude) shows a steady variation when moving from the metal-poor
(\feh=--3.0/--2.5) to the metal-rich (\feh=--0.5/0.0) regime. The smooth
transition in the peak of the period distribution as a function of the
metallicity strongly indicates that the long-standing problem of the Oosterhoff
dichotomy among Galactic globulars is the consequence of the lack of
metal-intermediate clusters hosting RRLs. We also found that the luminosity
amplitude, in contrast with period, does not show a solid correlation with
metallicity. This suggests that period-amplitude-metallicity relations should be
cautiously treated. 
\end{abstract}

\keywords{Stars: variables: RR Lyrae --- Galaxy: halo --- 
Techniques: spectroscopic}

%_____________________________________________________________________
\section{Introduction} \label{sec:intro}

The advent of space telescopes (HST, Kepler, {\it Gaia}) together with long term
photometric surveys (OGLEIV, VVV, ASAS, CATILINA) and high-resolution
multi-object spectrographs (GIRAFFE@VLT, GMOS@Gemini, AAOmega@AAT) at
ground-based 8-10m class telescopes are paving the way to a new golden age for
stellar evolution and resolved stellar populations. This means the opportunity
to estimate and to measure with unprecedented precision not only intrinsic
parameters such as stellar radius, effective temperature and stellar mass
\citep{pietrzynski13,pradamoroni12,marconi05}, but also the opportunity to
constrain the micro (atomic diffusion, opacity, equation of state) and the macro
(mixing, rotation, mass loss) physics adopted to construct evolutionary and
pulsation models \citep{salaris18}.   

In spite of this indisputable progress, there are several long-standing
astrophysical problems for which, after more than half a century of quantitative
astrophysics, we still lack an explanation based on plain physical arguments.
The so called {\em Oosterhoff dichotomy} is among the most appealing ones.
More than seventy years ago, \citet{oosterhoff39} recognised that RR Lyraes (RRLs) in
Galactic Globular Clusters (GGCs) can be split, according to the mean period of
the RRLs pulsating in the fundamental mode (RRab), in two different
groups: the Oosterhoff type I [OoI], with
$<$$P_{ab}$$>$$\sim$0.56~days, and the Oosterhoff type II [OoII], with longer
periods $<$$P_{ab}$$>$$\sim$0.66~days. The mean period of the RRLs pulsating
in the first overtone (RRc) displays a similar dichotomic distribution with
$<$$P_c$$>$$\sim$0.31~days and $<$$P_c$$>$$\sim$0.36~days in OoI and OoII
globulars, respectively. 
Subsequent spectroscopic investigations enriched the empirical scenario
demonstrating that OoI globulars are more metal-rich and cover a broad range in
metal abundances, while OoII globulars are more metal-poor stellar systems
\citep{arp55,kinman59}.  
Later on, it was also recognised that the population
ratio, i.e. the ratio between RRc and the total number of RRLs, is smaller in
OoI (N$_c$/N$_{tot}$$\approx$0.29) than in OoII (N$_c$/N$_{tot}$$\approx$0.44)
globulars \citep{stobie71,braga16,bono16}.

The literature concerning the Oosterhoff dichotomy is quite impressive. There
is no doubt that Allan Sandage provided in a series of papers covering half
century solid empirical evidence concerning the variation of the mean period in
field and cluster RRLs \citep[][and references therein]{sandage81,sandage81b,
sandage82,sandage90,sandage93,sandage06,sandage10}. This is the main reason
why the same problem is also quoted in the recent literature as the
Oosterhoff-Arp-Sandage period-shift effect \citep[][and references therein]{catelan09}. 
In this context it is worth mentioning the detailed
theoretical investigation provided by \citet{lee94} suggesting that a
difference in helium content ($\Delta$Y=0.03) could not explain the
observed variation in period, because the predicted variation in
period has an opposite sign. The same authors were more in favour of
a difference in absolute age of 1-2 Gyr between inner and outer halo
globular clusters to take account of the observed variation in period.
A difference in
luminosity between Oosterhoff I and II groups was also suggested by
\citet{lee99}. They investigated RRLs in M2 (OoII) and in M3 (OoI) and found
that the former sample was 0.2 magnitude brighter than the latter one. This
difference in luminosity was suggested to be caused by a difference in cluster
age \citep{lee90}. In particular, the RRLs in OoII clusters were considered
already evolved off the Zero-Age-Horizontal-Branch (ZAHB), while those in OoI
clusters were still near the ZAHB. Moreover, they also suggested, following
\citet{vandenbergh93a,vandenbergh93b}, there is a difference in kinematic
properties between OoI and OoII clusters. Indeed, the former ones appear to have
either vanishing or retrograde rotation, while the latter prograde rotation. On
the basis of these evidence they suggested that the OoII clusters formed in situ
in an earlier epoch, while the OoI clusters either formed later on or accreted.
The reader interested in a detailed discussion concerning theoretical and
empirical evidence concerning the Oosterhoff dichotomy at the of the last
century is referred to the review paper by \citet{caputo98}.

Evolutionary and pulsation prescriptions were taken into account by
\citet{castellani03} and they suggested that the difference between OoI and
OoII clusters could be explained as a consequence of a difference in the
topology of the RRL instability strip \citep{bono95}. On the basis of several
empirical evidence (the continuity of the mean fundamentalised period, the
period distribution in OoI and OoII clusters, the population ratio, the
difference between mean fundamental periods and fundamentalised periods) they
suggested that the so-called “OR" region\footnote{The region of the instability
strip in which the RRLs can pulsate either in the fundamental or in the first
overtone or in both of them \citep{bono94b}.} in OoI clusters is populated by
fundamental RRLs, while in OoII clusters is populated by first overtones. The
reader interested in a detailed discussion concerning the use of synthetic HB
models and their impact on the Oosterhoff dichotomy is referred to
\citet{cassisi04} and to \citet{catelan09}.

The possible occurrence of an Oosterhoff III group was also suggested by
\citet{pritzl03} to take account for the long mean fundamental period of RRLs in
two metal-rich clusters (NGC~6388, NGC~6441), but see also \citet{braga16}. The
empirical and theoretical scenario concerning the Oosterhoff dichotomy was
further enriched in a recent investigation by \citet{janglee15} in which the
authors suggested that the difference among OoI, OoII and OoIII clusters was a
consequence of multiple populations in Galactic globulars \citep{gratton04}. In
particular, they suggested that two/three different star formation episodes with
time delays ranging from $\sim$0.5 to $\sim$1.5~Gyr in inner and outer halo
clusters could explain the Oosterhoff-Arp-Sandage period-shift effect.

Large photometric surveys disclosed that Galactic field RRLs display a similar
dichotomy in the period distribution (\citealt{bono97a}; ASAS:
\citealt{pojmanski02}; LONEOS: \citealt{miceli08}; LINEAR:
\citealt{sesar13linear}).    
Oddly enough, Local Group galaxies (Draco, \citealt{kinemuchi08}; Ursa Minor,
\citealt{nemec88}; Carina, \citealt{coppola13}; Leo~I, \citealt{stetson14}) and
their globulars \citep{bono94} are characterised by mean fundamental periods
that fill the so-called \textit{"Oosterhoff gap"}, i.e. their mean periods range
from $\sim$0.58 to $\sim$0.62 days \citep{petroni04,catelan09}. The lack of
Galactic stellar systems with mean periods in the Oosterhoff gap indicates 
that the environment affects the Oosterhoff dichotomy \citep{coppola15,fiorentino15}. 

The analysis of this long-standing astrophysical problem was hampered by several
empirical biases.

\textit{a)} -- The number of GGCs with a sizeable (more than
three dozen) sample of RRLs is limited to 18 out of $\approx$100 globulars hosting
RRLs \citep{clement01}. This problem becomes even more severe for Ultra Faint
Dwarf galaxies in which the RRL sample never exceeds a dozen
\citep{dallora12,fiorentino15}.   

\textit{b)} -- Although, cluster RRLs have been investigated for
more than one century \citep{bailey1902}, the current samples are far from being
complete. This limitation applies to objects centrally located and to low
amplitude variables. The same problem applies to nearby dwarf galaxies 
due to the lack of a full spatial coverage.

\textit{c)} -- There is mounting empirical evidence that
old- and intermediate-age stellar populations in nearby dwarf galaxies display
different metallicity distributions \citep{fabrizio15}. This means that RRLs in
dwarf galaxies might be the progeny of stellar populations characterised by a
broader age and/or metallicity distribution \citep{martinez15} when 
compared with cluster RRLs. The same outcome applies to RRLs in $\omega$
Centauri, the most massive GGC \citep{braga16}.     

\textit{d)} -- The Bailey diagram (period \textit{vs} luminosity
amplitude) is a solid diagnostic, since it is---together with the period 
distribution---independent of distance and reddening. To
constrain the RRL intrinsic properties, \citet{stetson14} and
\citet{fiorentino15} found that the High Amplitude Short Period (HASP,
$P$$<$0.48~days, $A_V$$>$0.75~mag) variables are not present in dwarf
spheroidals, with the exception of Sagittarius. Detailed investigation among 
clusters with sizeable sample of RRLs indicate that HASP are only present 
in systems that are more metal-rich than \feh=--1.5 \citep{monelli17}.

In the following, we will focus our attention on the pulsation properties 
of halo RRLs as a function of the chemical composition. 
The structure of the paper is as follows. In Section 2 we introduce the
photometric data sets we adopted to build up the master catalogue of candidate
field RRLs. Special attention is given to the cross-match between the RRLs
catalogues available in the literature and the {\it Gaia} DR2 catalogue. In this
section we also mention the criteria we adopted to select candidate Halo RRLs
and the approach adopted to identify fundamental and first overtone RRLs.
In Section 3 we introduce the spectroscopic data sets we adopted to build up the
RRL spectroscopic catalogue. In this section we also describe the approach adopted
to retrieve the SDSS-SEGUE medium-resolution spectra and the variant of the
\deltaS\ method adopted to estimate the metallicity of individual RRLs.
Moreover, we also discuss the spectroscopic data sets available in the
literature.
Section 4 deals with the strategy adopted to calibrate and to validate the
metallicity scale based on the \deltaS\ method. In particular, we focus our
attention on V~Ind, a fundamental field RRL, for which we have X-shooter spectra
covering the entire pulsation cycle. 
In Section 5 we discuss the metallicity distribution of fundamental RRLs and the
comparison with similar estimates available in the literature.
Section 6 deals with the fine structure of the Bailey diagram, and in particular, 
its dependence on the metal content. In this
section we also introduce some long-standing open problems connected with the
Oosterhoff dichotomy and provide new analytical period-metallicity and 
period-amplitude relations. 
In Section 7 we briefly discuss the use of the period distribution and of the
amplitude distribution to constrain the key properties of the underlying stellar
populations. We focus our attention on the RRLs in the Bulge, in Galactic
globular clusters, in Magellanic Clouds and in nearby dwarf galaxies. 
Finally, Section 8 gives a summary of the current results together with a few
remarks concerning the future developments of this project.

\begin{figure*} %%%%%%%
\centering
\includegraphics[width=1.8\columnwidth]{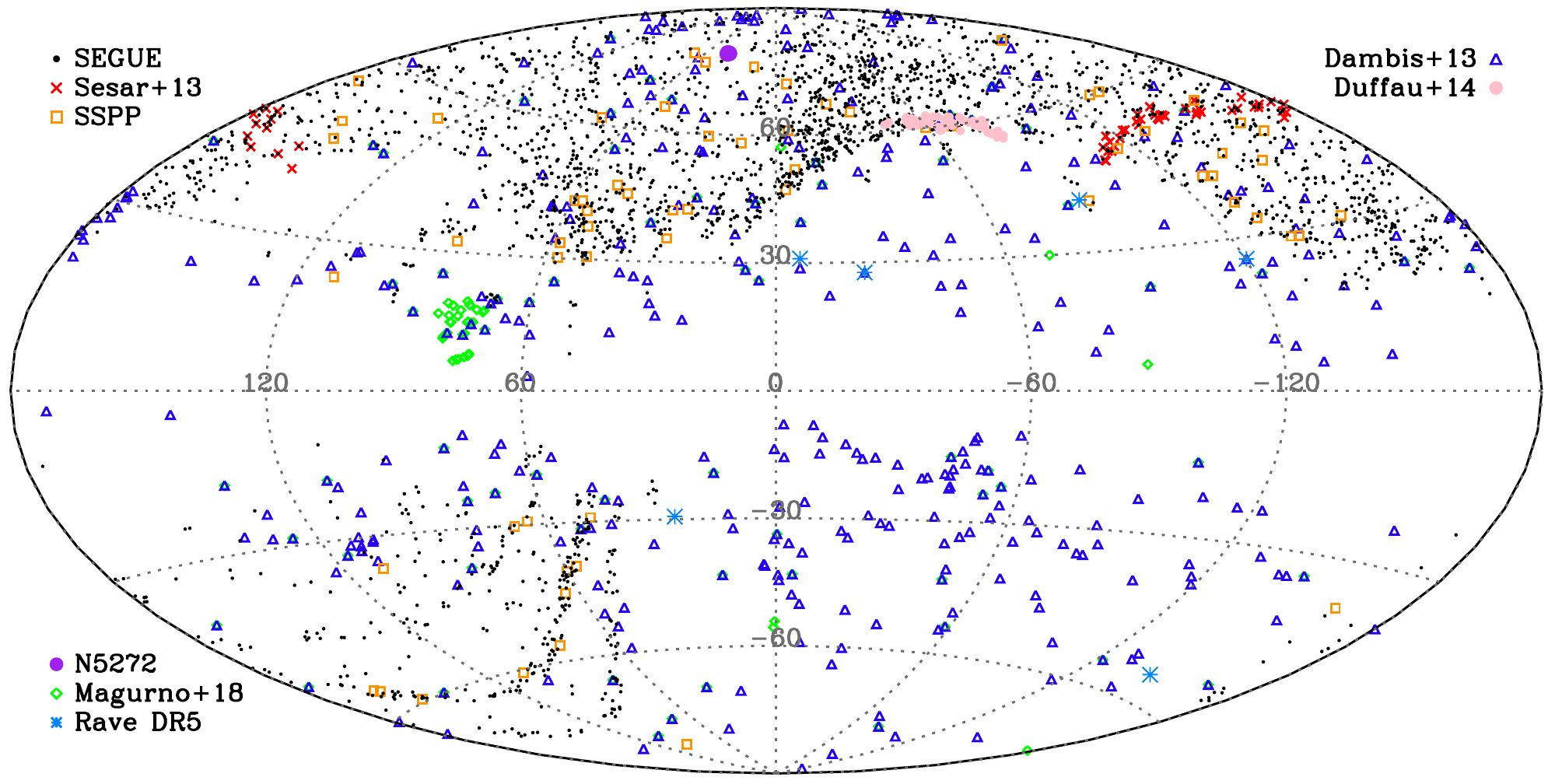} 
\caption{Distribution in Galactic coordinates of the RRL spectroscopic sample
(2,903 stars). The black circles show the RRLs with iron abundances based on \deltaS\ method on low-resolution SDSS-SEGUE spectra (2,382, SEGUE), while the red crosses display RRLs with iron abundances from \citet[50]{sesar13}. Orange squares show RRLs with iron abundances based on SDSS-SSPP indicators (65, SSPP). The blue triangles and the pink circles display RRLs with metallicities from \citet[360]{dambis13} and \citet[57]{duffau14}. Purple circles are used for the five variables in NGC~5272. Green diamonds show the distribution of the RRLs with iron abundances based on high spectral-resolution spectra \citep[104]{magurno18}, while the cyan asterisks refer to iron abundances from Rave DR5 (6).}\label{fig:aitoff}
\end{figure*} %%%%%%%

%_____________________________________________________________________
\section{Photometric data sets} \label{dataset}
\subsection{Photometric catalogue}\label{sec:photoCat}

To provide firm constraints on the metallicity distribution of the Galactic halo
we used different photometric and spectroscopic catalogues available in the
literature together with the exquisite data provided by the ESA mission {\it
Gaia} \citep{gaia18}. A detailed description of the construction of such
catalogue is provided in two companion papers (Marinoni et al. 2019, in
preparation; Bono et al. 2019, in preparation). Here we briefly summarise the
main steps of this process.

We started from the following list of published optical RRL catalogues 
and surveys:
\citet{dambis13}, %table 2
CATALINA \citep{torrealba15,drake13a,drake13b,drake14},
  %Torrealba 2015 2015MNRAS.446.2251T table 2 (CATALINA)
  %Drake 2013a 2013ApJ...763...32D table 1(CATALINA)
  %Drake 2013b 2013ApJ...765..154D table 1-2 and 4(CATALINA)
  %Drake 2014  2014ApJS..213....9D table 3(CATALINA)
LINEAR \citep{sesar13linear}, %table 1 (LINEAR)
LONEOS-I \citep{miceli08}, %table 1 (LONEOS-I)
NSVS \citep{hoffman09}, %table 2 (NSVS)
ROTSE~I \citep{akerlof00}, %table 1 (ROTSE I)
QUEST~I \citep{vivas04}, %table 2 (QUEST I)
ASAS \citep{pojmanski02}, %table asas-v.dat (ASAS)
ASAS-SN \citep{shappee14,jayasinghe18}, % (ASAS-SN)
\citet{magurno18}. % table 10

We build up a single catalogue containing all entries of the quoted
literature samples ($\sim$42,000), with a particular care to recognise RRLs which were listed
in more than one catalogue. We used the algorithm described in \citet{marrese19}
for sparse catalogues to cross-match the literature RRLs with {\it Gaia} DR2
data, keeping only those stars with a {\it Gaia} counterpart. Moreover, we added
the new RRLs detected by {\it Gaia} \citep{clementini19}, which were not
included in the literature. The final catalogue includes more than 150,000
candidate RRLs. In order to collect multi-band magnitudes, we used the powerful
results of the official {\it Gaia} cross-match \citep{marrese17,marrese19}. In
particular, we were able to recover near-infrared (NIR: \jj, \hh, \kk)
magnitudes from 2MASS PSC \citep{skrutskie06} and VHS DR3 \citep{mcmahon13},
mid-infrared (MIR: $W1$, $W2$) magnitudes from allWISE \citep{wright10,cutri13}
and optical (\usdss, \gsdss, \rsdss, \isdss, \zsdss) magnitudes from SDSS DR9
\citep{ahn12,alam15}. Aiming at a wider wavelength coverage, we
also performed, by using the algorithm developed for the large dense surveys,
the cross-match of {\it Gaia} DR2 with ultra-violet catalogue (UV: \fuv, \nuv) from GALEX
GUVcat\_AIS \citep{bianchi17}. This means that we build up an RRL photometric
catalogue including magnitudes from the UV to the MIR.  

%_____________________________________________________________________
\subsection{Selection of field Halo RRLs}

To improve the selection of field Halo RRLs we applied several selection
criteria discussed in the following. It is worth mentioning that they are
conservative, i.e. we preferred to possibly lose some candidates, but to avoid
spurious contaminations with false identification, and/or blended targets and/or
Thin Disk variables such as High Amplitude $\delta$ Scuti.
We, also, provided a preliminary estimate of the individual distances by
using predicted optical, NIR and MIR Period-Luminosity relations provided by 
\citet{marconi15,marconi18}. The individual distances were estimated by using 
apparent MIR and NIR mean magnitudes from allWISE and 2MASS/VHS. 
Note that in this preliminary step we neglected distances based on 
{\it Gaia} \citep{bailerjones18} because the current RRL sample approaches 
the outermost Halo regions ($\sim$100~kpc).
The distances of the RRLs, for which NIR/MIR measurements were not available, were
derived by adopting \rsdss, \isdss, \zsdss-band photometry from the SDSS. The distance of
the RRLs lacking both MIR/NIR and SDSS photometry was estimated by using the
mean \G, \BP\ and \RP\ magnitudes provided by {\it Gaia}. The mean of the individual
\G-band measurements was transformed into a mean \rr-band magnitude by using the
transformations provided by \citet{evans18}.
Finally, the distances of RRLs, for which at least one of the three quoted {\it Gaia} 
magnitudes was not available, was evaluated by using the canonical 
visual magnitude-metallicity relation ($M_V$ \textit{vs} \feh) recently provided by 
\citet{marconi18}. 
For these variables the mean visual magnitude was retrieved from the literature 
and we adopted a mean Halo metallicity of \feh=--1.65 \citep{layden93}. 
The reader interested in a more detailed discussion concerning the Halo 
metallicity distribution is referred to Sect.~\ref{sec:catalogue}.  
The MIR/NIR and optical apparent mean magnitudes were un-reddened by using the
\ebv\ values from \citet{schlafly11}, which is the recalibrated extinction map 
of \citet{schlegel98}, and the \citet{cardelli89} reddening law.

{\em Extended Sources} -- We removed the objects flagged as "extended" in the
2MASS PSC and allWISE catalogues, by using \texttt{extKey} ($\neq$NULL) 
and \texttt{extFlag} ($>$1) columns respectively.

{\em Position and reddening} -- In order to avoid the Galactic plane and/or
highly reddened areas, we decided to remove the candidate RRLs located either
within $\pm$2.5 degrees from the Galactic plane or with a reddening
\ebv$\ge$2mag.

{\em Spatial overdensities} -- The distribution of the entire catalogue in
Galactic coordinates (X, Y, Z) shows several well-defined overdensities
associated either to nearby dwarf galaxies (Magellanic Clouds, Ursa Minor,
Draco, Sculptor, Fornax, Carina) or to a globular (NGC~2419) or to the
Sagittarius stream \citep{majewski03}. They were flagged and the stars belonging
to dwarf galaxies or to the globular cluster were removed from the master
catalogue.
Note that we forced the inclusion of five cluster RRLs belonging to NGC~5272
to increase the sample of spectroscopic standards adopted for
calibrating the \deltaS\ metallicity scale (see Sect.~\ref{sec:otherData}). 

{\em Spectral Energy Distribution} -- To further improve the selection of
candidate Halo RRLs we also used their Spectral Energy Distribution (SED). The
current master catalogue includes multi-band UV (GALEX), optical 
(SDSS; {\it Gaia}; literature: \vv, \ii), NIR (2MASS; VHS) and MIR (allWISE)
magnitudes. We took advantage of these independent measurements to estimate 
several un-reddened mean colours ($m_G - m_\lambda$)$_0$
as a function of $\lambda$. 
On the basis of the RRLs already known in the literature ($\sim$42,000) we
defined in the colour-$\lambda$ plane a template for the expected RRL colours.
We performed an analytical fit of the colour variation and excluded those
objects located outside 1$\sigma$ from the analytical fit.   

{\em Galactocentric distance} -- We removed from the sample the candidate RRLs
located closer than 4.5~kpc from the Galactic Center. This is a conservative
threshold which allows us to neglect \textit{bona-fide} Galactic Bulge RRLs 
\citep{pietrukowicz15,valenti18,zoccali18}.

After the last selection criterium, we obtained a cleaned master catalogue of
candidate Halo RRLs for which, together with the quoted parameters, we also have
an estimate of their pulsation period and visual amplitude. For more than 90\%\
of the sample, we have adopted Amp(\G) from Gaia, while for the remaining 10\%,
we have adopted literature data from the surveys introduced in
Sect.\ref{sec:photoCat}. The light curves of the latter sample were visually
inspected and, for a fraction of them, we performed a new estimate of the
luminosity amplitudes by using the original time series. The luminosity
amplitude in Amp(\G) was transformed into Amp(\vv) by using the Eqn.2 from
\citet{clementini19}. In passing we also note that the current luminosity
amplitudes are minimally affected by Blazhko modulations, since the cadence and
the time interval covered by the adopted long-term photometric surveys cover
tens of amplitude modulation cycles.
Moreover, to provide a homogenous mode classification we adopted the
period-amplitude criterium suggested by \citet{clementini19}:
\begin{equation}
\frac{2.08-{\rm Amp}(G)}{3.5}<P ({\rm days})
\label{eqn:rrab}\end{equation}
where ${\rm Amp}(G)=[{\rm Amp}(V)-0.013]/1.081$~mag. Note that we only included
RRL candidates with pulsation periods ranging from 0.2 to 1.0 day. 

Finally, we neglected both first overtone and mixed mode RRLs by using Eqn.~\ref{eqn:rrab}
and we ended up with a sample of 44,822 RRab.

%_____________________________________________________________________
\section{Spectroscopy data sets}\label{spec}

The photometric RR Lyrae data sets were complemented with spectroscopic data
sets based either on high- or on medium- or on low-resolution spectra. As a
whole, we ended up with a sample of 2,903 RRab variables with an iron abundance
estimate based on a spectroscopic measurement. Note that in the following we 
are only dealing with RRab variables, because the spectroscopic calibration 
adopted for the bulk of the data
was devised for this group of variables (see Sect.~\ref{sec:deltas}). 
The first overtone RRLs will be addressed in a forthcoming paper 
(Fabrizio et al. 2019, in preparation).  

In the following we discuss the different spectroscopic data sets 
together with the approach adopted to calibrate them on a homogenous 
metallicity scale. Moreover, we also introduce the approach adopted 
to validate the spectroscopic diagnostics we are using to estimate 
iron abundances.

\begin{figure}%%%%%%%
\centering
\includegraphics[width=\columnwidth]{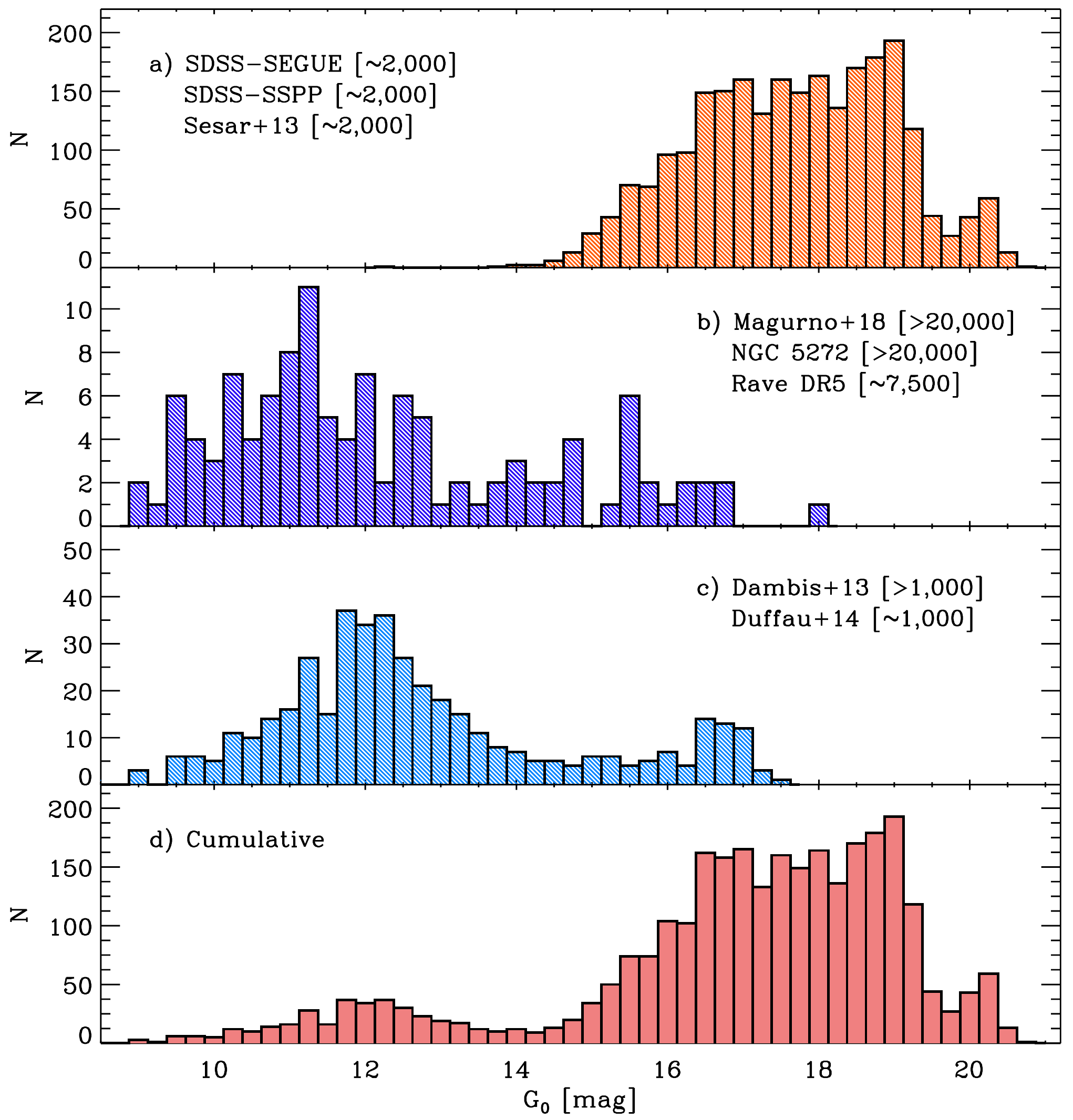} 
\caption{From top to bottom, un-reddened \G\ magnitude distributions of 
the RRL samples with spectroscopic measurements.
The values enclosed in square brackets refer to the spectral resolution of the
various samples.}\label{fig:distr_magn}
\end{figure} %%%%%%%

%_____________________________________________________________________
\subsection{SDSS-SEGUE data}
\label{sec:segueData}

\begin{figure*} %%%%%%%
\centering
\includegraphics[width=1.8\columnwidth]{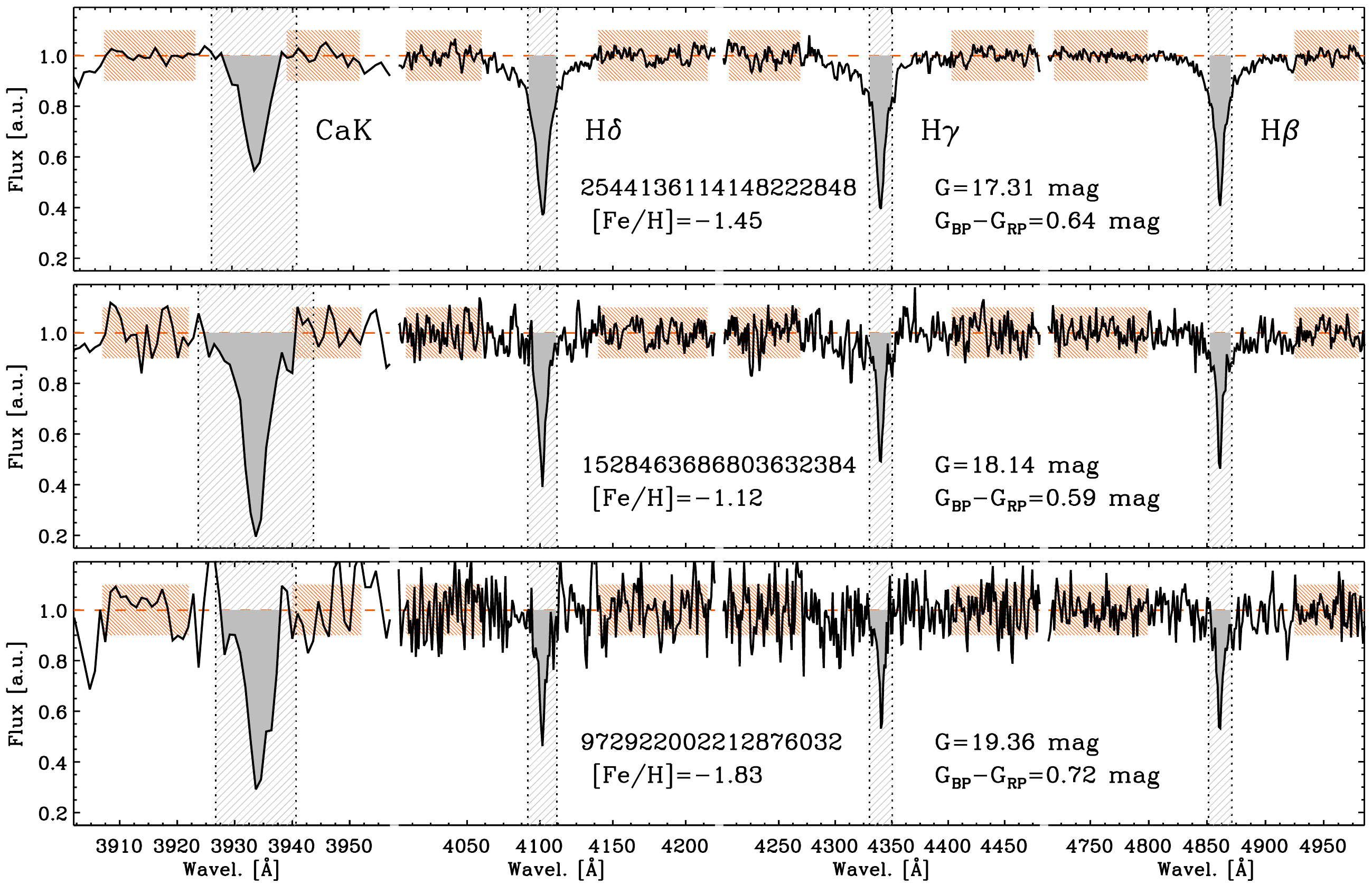} 
\caption{Normalised SEGUE spectra for three field RR Lyrae. The
hatched orange regions outline the wavelength range used to estimate the
continuum mean flux (red dashed line), while the hatched light grey regions 
and the dotted vertical lines display the wavelength interval in which the 
equivalent width is measured. The dark grey regions display the EWs for the 
four spectroscopic diagnostics: \caiik, \hdelta, \hgamma\ and \hbeta.}
\label{fig:spectra}
\end{figure*} %%%%%%%

We focussed our attention on the survey "Sloan Extension for Galactic
Exploration and Understanding" (SEGUE, \citealt{yanny09}) based on medium
resolution (R$\sim$2,000) spectra collected with the Sloan
Digital Sky Survey DR12 (SDSS, \citealt{alam15}). The photometric and the
spectroscopic data collected in this survey are publicly available from the SDSS
Science Archive Server (SAS)\footnote{\url{https://dr14.sdss.org/home}}. The
initial step was to download all the available SEGUE spectra for the RRLs in our
photometric catalogue. The search was based on the \texttt{bestObjID} from the
\texttt{SpecObjAll} table, and we ended up with 2,382 RRab variables for which
are available the SEGUE "lite" spectra, i.e. the co-added spectra including 
up to 38 individual measurements. The sky distribution in Galactic coordinates of the SEGUE
sample is shown in Fig.~\ref{fig:aitoff} (black symbols).

Fig.~\ref{fig:distr_magn} shows the un-reddened \G-band magnitude distribution
for different sample of RR Lyrae stars, in particular the SEGUE sample is 
displayed in panel a) (see also Sect.~\ref{sec:otherData}). 
The individual reddening values were extracted from the
\citet{schlegel98} dust maps and the updated reddening coefficients from
\citet{schlafly11}, while the extinction in \G\ band was calculated with the
\citet{casagrande18} relation. 
The key feature of the SEGUE spectra is that they cover a
broad spectral range, namely from 3800 to 9200\AA\ and the majority of the
spectra have a mean signal-to-noise ratio larger than $\sim$20 in the 
blue region (3900-4900\AA). This is the main reason why we decide to use 
the \deltaS\ method introduced half a century ago by G.~W. Preston to 
estimate the iron content of RRLs (see Sect.~\ref{sec:deltas}).

%_______________________________________________________________________________
\subsubsection{Metal abundances based on the \deltaS\ method} \label{sec:deltas}

We derived abundances using a variation of the \deltaS\ method originally
introduced by \citet{preston59}. In particular, we are following the same
approach developed by \citet{layden94} which is based on the comparison of
pseudo-equivalent width of the \caiik\ line, W(K), and of the mean
pseudo-equivalent width of hydrogen lines \hdelta, \hgamma\ and \hbeta, W(H).
The pseudo-equivalent widths (hereafter, EW) were measured on SEGUE spectra by
using an
\texttt{IDL}\footnote{\url{https://www.harrisgeospatial.com/Software-Technology/
IDL}} version of the original \texttt{EWIMH}
program\footnote{\url{http://physics.bgsu.edu/~layden/ASTRO/DATA/EXPORT/EWIMH/
ewimh.htm}} written by one of us (A. Layden). The algorithm defines, for each spectral
feature, a pseudo-continuum level as a straight line (dashed red line in
Fig.~\ref{fig:spectra}) between the mean intensity and the mean wavelength
points of two continuum bands (see table~5 in \citealt{layden94} and the orange
hatched areas in Fig.~\ref{fig:spectra}). 

The EW (dark grey area showed in Fig.~\ref{fig:spectra}) is defined as the area
enclosed by the limits in wavelength of the specific spectral feature (vertical
light grey hatched area) and the pseudo-continuum of the spectrum. This area is
then divided by the mean height of the continuum inside the specific spectral
feature. The three panels of Fig.~\ref{fig:spectra} display the details of the
measurements for three targets with different magnitudes, colours and
metallicities (see labeled values). 

A crucial issue in the use of the \deltaS\ method is the calibration of the
measured EWs onto a "standard sample" of EWs. The list of the 17 standard stars
is given in Table~6 of \citet{layden94}. Unfortunately, there is no overlap
between the SEGUE survey and the set of spectroscopic "standards" adopted by
Layden. This means that we cannot directly use the relations defined by
\citet{layden94} to derive the iron abundance. Fortunately enough, in a recent
investigation, one of us \citep{sesar13} provided an independent calibration of
the \deltaS\ method to investigate the metallicity distribution of RRLs in the
Orphan Stream. They collected low-resolution spectra (R$\sim$1350) with the
Double Spectrograph (DBSP, \citealt{oke82}) available at the Palomar 5.1m
telescope for 50 Orphan Stream RRLs. Moreover, they also observed eight out of
the 17 standard stars and provided four linear relations between the EWs
measured on DBSP spectra and those based on the Layden's spectroscopic standards
(see equations 9-12 in \citealt{sesar13}). 

\begin{figure} %%%%%%%
\centering
\includegraphics[width=1\columnwidth]{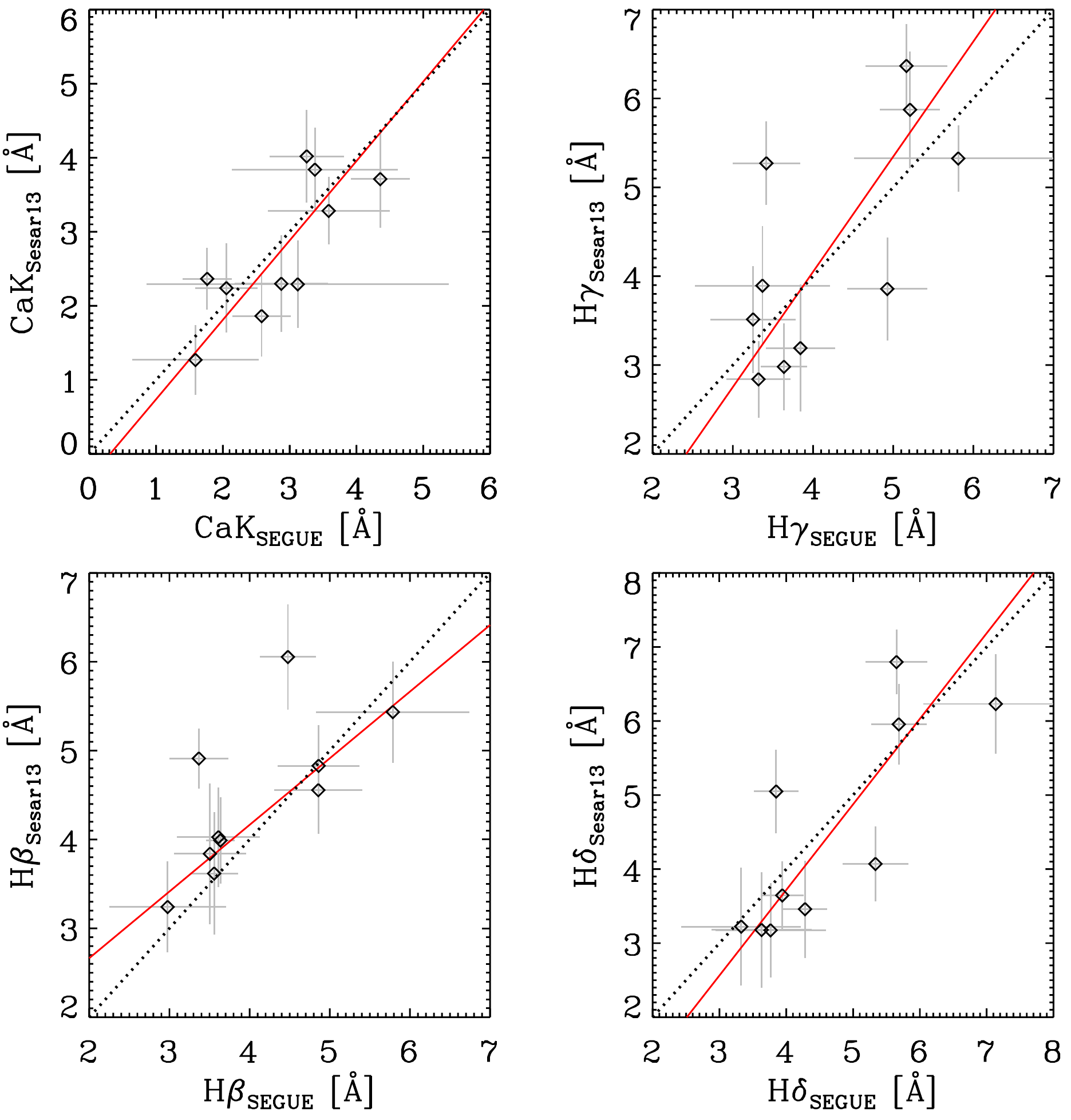} 
\caption{Comparison between the equivalent widths measured on the  
\citet{sesar13} spectra and those based on the SEGUE spectra degraded 
to the DBSP resolution (R$\sim$1300) for the ten RRLs in common. 
The dotted line shows the bisector of the plane.
The linear fits used to transform the EWs based on SEGUE spectra 
into the Sesar's equivalent width system is plotted as a red line.}
\label{fig:EWcfr}
\end{figure} %%%%%%%

The Orphan Stream spectroscopic data set and the SEGUE data set have 27 RRLs in
common. Among them, ten have a sufficient signal-to-noise ratio to calibrate
the EWs measured on SEGUE spectra onto the EWs measured on DBSP spectra. More
specifically, we measured the EWs on both DBSP
and on SEGUE spectra (degraded to the DBSP spectral resolution). 
The individual measurements concerning the \caiik\ line and the three hydrogen
lines are plotted in Fig.~\ref{fig:EWcfr} and show, within the errors, a
linear trend over a broad range of EWs. We also performed four linear fits to
transform the current EW measurements into the EW system defined by
\citet[][see red lines]{sesar13}. The linear relations are the following: 
\begin{eqnarray}
{\rm CaK_{Sesar13}} &=& 1.07\cdot {\rm CaK_{SEGUE} - 0.34} \\
{\rm H\beta_{Sesar13}} &=& 0.75\cdot {\rm H\beta_{SEGUE} + 1.17} \\
{\rm H\gamma_{Sesar13}} &=& 1.30\cdot {\rm H\gamma_{SEGUE} - 1.14} \\
{\rm H\delta_{Sesar13}} &=& 1.16\cdot {\rm H\delta_{SEGUE} - 0.90}
\end{eqnarray}

\begin{figure} %%%%%%%
\centering
\includegraphics[width=1\columnwidth]{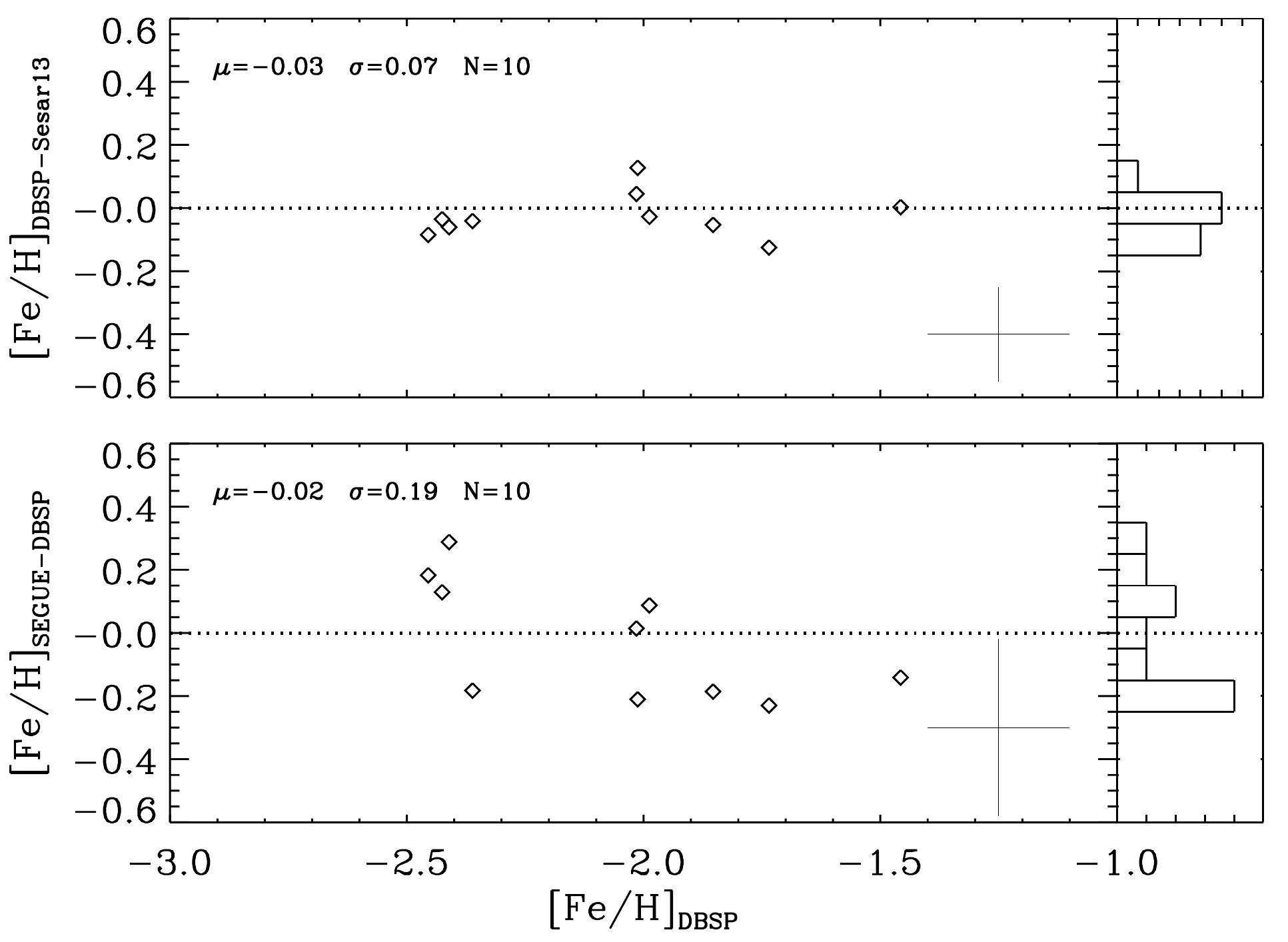} 
\caption{\textit{Top}: Difference between the iron abundances estimated by us on the DBSP spectra
 and those provided by \citet{sesar13}.
\textit{Bottom:} Difference between the 
iron abundances based on the \deltaS\ method applied to the re-binned 
SEGUE spectra and the iron abundance we estimated with the \deltaS\ 
method applied to \citet{sesar13} DBSP spectra, for ten stars in common.
The mean and the standard deviation of the differences are also labelled.}
\label{fig:sesar}
\end{figure} %%%%%%%
 
Finally, the EWs in the system defined by \citet{sesar13} were transformed into
the system defined by \citet{layden94} standard stars. Moreover, following
\citet{layden94}, the \caiik\ EWs were also corrected for interstellar \caii\
absorption using the \citet{beers90} model:
\begin{equation}
   W(K_0)=W(K)-W_{\rm max} (1 - e^{-|z|/h})/\sin{|b|},
\end{equation}
where $W_{\rm max}=0.192$\AA, $h=1.081$ kpc, $b$ is the Galactic latitude, and
$z$ is the height above the Galactic plane in kpc.

The iron abundances were evaluated by inverting equation 7 
of \citet{layden94}:
\begin{equation}
   \feh = \frac{W(K_0) - a - bW(H)}{c + dW(H)},
\end{equation}
where $a=13.858$, $b=-1.185$, $c=4.228$ and $d=-0.32$. To validate the current
approach, we compared our iron abundances with those provided by
\citet{sesar13}. The top panel of Fig.~\ref{fig:sesar} shows the comparison
between the \feh\ measured by us on the DBSP spectra and the \feh\ abundances
measured by \citet{sesar13}. We found a very good agreement, and indeed, the
mean difference is minimal ($-0.03$~dex) and the dispersion is negligible
($0.07$~dex). The bottom panel of the same Fig.~\ref{fig:sesar} shows a
similar comparison, but between the \feh\ abundances based on SEGUE and on DBSP
spectra. We found again a good agreement between the two data sets, and indeed
the mean difference is minimal ($-0.02$~dex) and the dispersion is smaller than
0.2~dex. These results further support the approach we devised to calibrate the
\feh\ abundance onto those provided by \citet{sesar13}, and subsequently onto
the Layden's metallicity scale \citep{layden94}, which, in turn, is rooted onto
the \citet{zinnwest84} globular cluster metallicity scale.

%_____________________________________________________________________
\subsection{Spectroscopic data sets available in literature}\label{sec:otherData}

In order to validate and to enlarge the SDSS-SEGUE dataset, we also included
the large sample of iron abundances collected by \citet[][Tab.10]{magurno18} and
based on high-resolution spectra (R$>$20,000). The whole dataset was scaled to
the \citet{asplund09} Solar reference. The entire sample includes 134 objects,
but we only took into account fundamental RRLs (104). Note that this sample
mainly includes bright nearby RRLs, and indeed the limiting magnitude is
\G$\sim$17 mag.

To increase the spatial distribution and the size of high-resolution sample, the
quoted data set was complemented with the iron abundances retrieved from the
Radial Velocity Experiment DR5 (RAVE, \citealt{kunder17,casey17}). The iron
abundances for six RRLs are based on spectra covering the Ca-triplet region
(8410--8795\AA) with a spectral resolution R$\sim$7,500. Furthermore, the sample
was complemented with five cluster RRLs belonging to NGC~5272. We adopted the
RRLs listed in \citet{clement01}, and the cluster iron abundances provided by
\citet{harris10}. The three data sets defining the high-resolution (HR) sample
were scaled to the same \citet{asplund09} Solar reference. Their \G$_0$-band
magnitude distribution is shown in the panel b) of Fig.~\ref{fig:distr_magn}.

%%%%
The literature sample was also complemented with the iron abundances collected
by \citet{dambis13}, based on a mix of low-, medium- and high-resolution
spectra. This data set includes 402 RRLs and among them 360 were included in 
the current spectroscopic catalogue. The bulk of this data set comes either from 
the \deltaS\ measurements provided by \citet{layden94}, by \citet{fernley98} 
and by \citet{kinman07}. Panel c) of Fig.~\ref{fig:distr_magn} shows the 
magnitude distribution of this data set.

Furthermore, we complemented the literature sample by including the
metallicities of RRLs, based on \deltaS\ method, identified by the QUEST survey
and published by \citet{duffau14}. This data set is based on a mix of low- and
medium-resolution spectra. This sample includes 82 RRLs and among them 57 are RRab 
variables belong to the current spectroscopic catalogue. Its magnitude distribution is shown in
the panel c) of Fig.~\ref{fig:distr_magn}, mainly defining the tail between
\G$_0$$\sim$16 and 18~mag.

%%%%
Moreover, the SEGUE survey also provides an independent estimate of the iron
abundance by using their Stellar Parameter Pipeline (SSPP, \citealt{lee08}). The
SSPP uses multiple techniques to measure the radial velocities, to estimate the
fundamental stellar parameters (effective temperature, surface gravity) and to
determine the iron abundance \citep{lee08b,allende08}. In this context, it is worth
mentioning that the iron abundances provided by SSPP are based on twelve
independent spectroscopic diagnostics. The pipeline gives a mean best value
(\texttt{FEHADOP}) together with its uncertainty. These iron estimates define
the SDSS-SSPP sample and among them 65 were included in the current spectroscopic
catalogue. The cumulative magnitude distribution of the entire spectroscopic
catalogue is shown in panel d) of Fig.~\ref{fig:distr_magn}.

%_____________________________________________________________________
\section{Calibration and validation of the spectroscopic data set}
%_____________________________________________________________________
\subsection{Spectroscopic Calibration} \label{sec:calib}

\begin{figure} %%%%%%%
\centering
\includegraphics[width=\columnwidth]{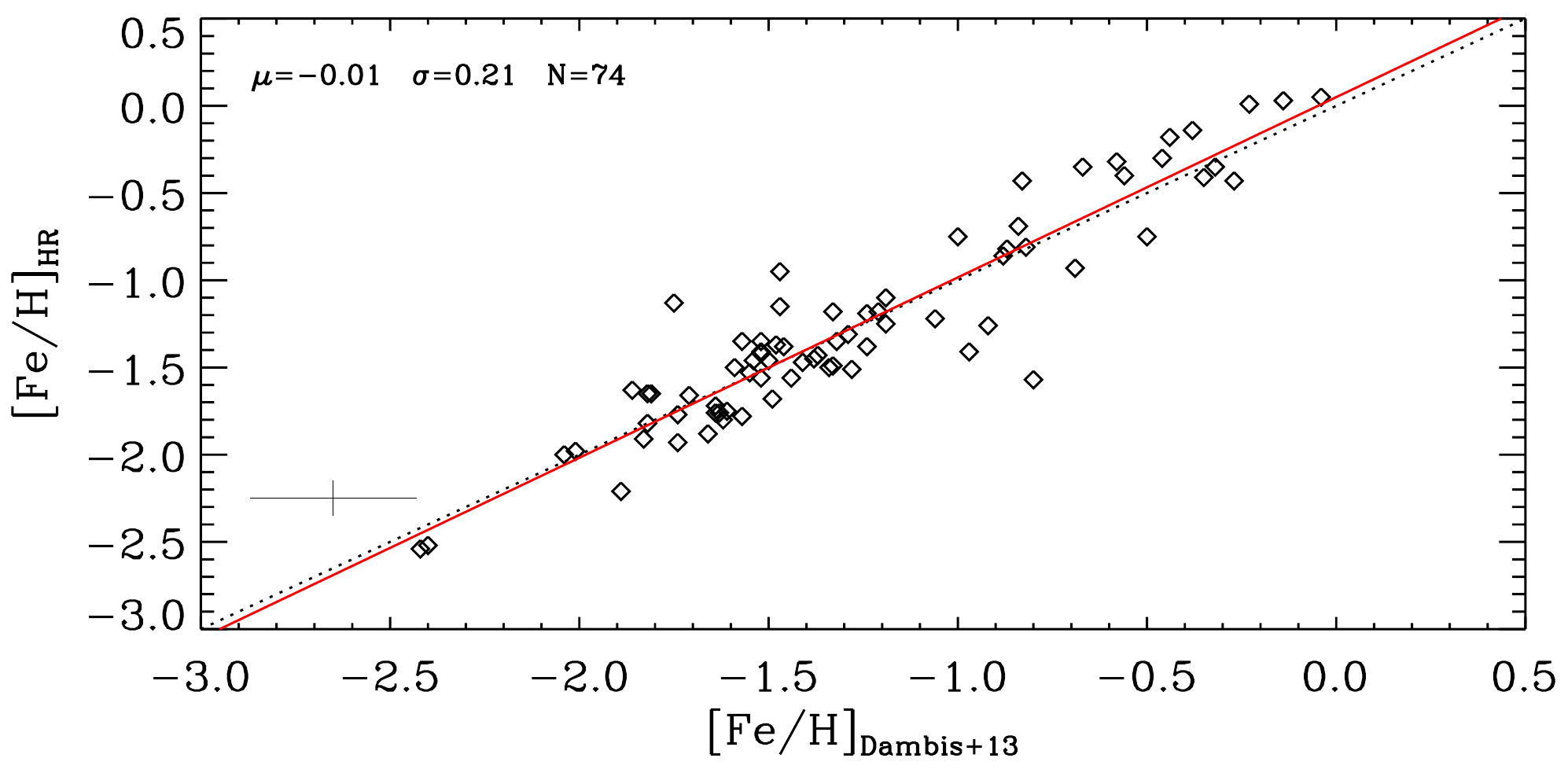} 
\caption{Calibration of \citet{dambis13} iron abundances with the 
iron abundances based on high-resolution spectra collected by 
\citet{magurno18}. The linear fit used to transform 
\citet{dambis13} iron abundances into the HR sample is 
plotted as a red line. The dotted line shows the bisector of the plane}\label{fig:hr}
\end{figure} %%%%%%%

To provide a homogenous metallicity scale for the different
spectroscopic data sets discussed in the previous section we took into account
stars in common between the HR sample (pivot sample), the medium- and
low-resolution data sets. We found 74 RRLs in common between the HR sample
and the \citet{dambis13} sample. Data plotted in Fig.~\ref{fig:hr} show that
the two data sets agree quite well, and indeed, the mean difference is minimal
($-0.01$~dex), while the standard deviation is $0.21$~dex. The dispersion is
mainly a consequence of the intrinsic errors of the two data sets (see error
bars in the bottom left corner). Note that the 74 RRLs in common cover a wide
range in \feh\ abundances (more than 2~dex) and we found evidence of a mild
drift when moving from the metal-poor to the metal-rich regime. We performed a
linear regression and we found the following linear relation:
\begin{equation}
   {\rm [Fe/H]_{HR}}=0.05+1.03 \cdot {\rm [Fe/H]_{Dambis+13}}
\end{equation}
to move the Dambis iron abundances into the HR metallicity scale. 

\begin{figure} %%%%%%%
\centering
\includegraphics[width=\columnwidth]{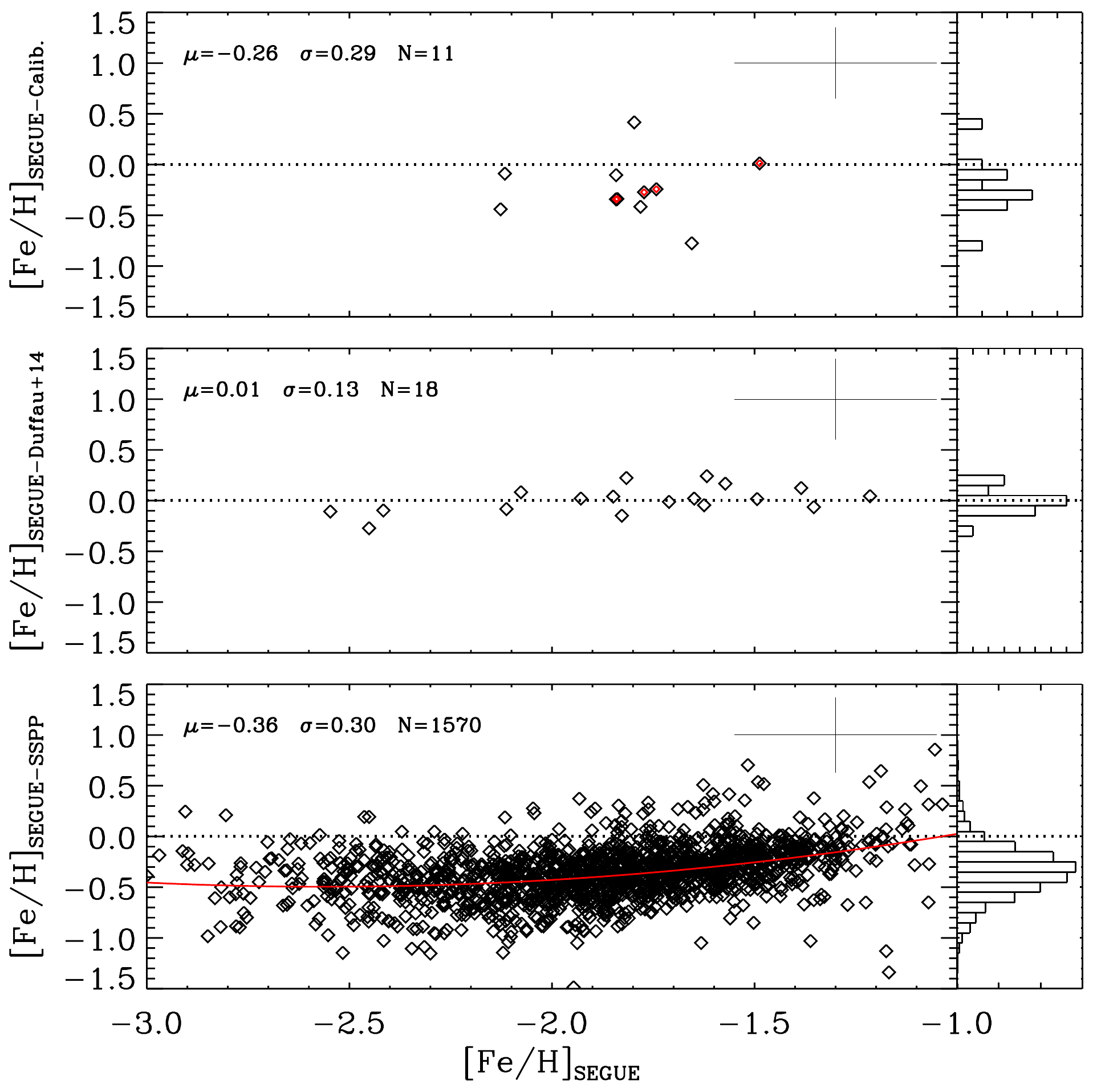} 
\caption{\textit{Top:} Comparison between the iron abundances based on the
current \deltaS\ method and those based on the calibration sample (see
Sect.~\ref{sec:calib}). The red diamonds mark the Globular Cluster RRLs. 
\textit{Middle:} Comparison between the iron abundances based on the
current \deltaS\ method and those based on the \citet{duffau14} sample.
\textit{Bottom:} Comparison between the iron abundances based on the \deltaS\ 
method and those based on the SDSS-SSPP method.
The quadratic relation to transform the SDSS-SSPP into the current 
metallicity scale is also plotted as a red line.
The mean and the standard deviation of the 
difference are also labelled together with the sample size.}
\label{fig:calibration}
\end{figure} %%%%%%%

The HR sample was joined with the Dambis sample, defining a new data 
set of 401 RRLs as the "calibration sample". The iron
abundances of the calibration sample were compared with values based on the
\deltaS\ method we applied to the SEGUE spectra. The number of RRLs in common is
eleven and the top panel of Fig.~\ref{fig:calibration} shows the comparison.
We found a systematic offset of $-0.26$~dex (with a standard deviation of
0.29~dex) and it was applied to the iron abundances based on the \deltaS\ method.  

The middle panel of Fig.~\ref{fig:calibration} shows the difference of \feh\
based on \deltaS\ method between the SEGUE and \citet{duffau14} samples. The 
two samples have 18 RRab variables in common and the mean difference in 
metallicity is vanishing (0.01~dex) with a dispersion of 0.13~dex. This result 
allows us to treat the \citet{duffau14} sample as the SEGUE one, by applying 
the same metallicity scale.

Finally, we calibrated the iron abundances based on the SDSS-SSPP metallicity
determinations into the same HR metallicity scale. 
The number of RRLs in common among the HR sample plus our \deltaS\ sample and
SDSS-SSPP is larger than 1,500 objects.
Data plotted in the bottom panel of
Fig.~\ref{fig:calibration} shows that the mean difference is $-0.36$~dex. Note
that the current finding agrees with a similar result (difference equal to 
$-0.36$~dex) obtained by \citet{sesar13} by using an independent spectroscopic 
data set. The dispersion we found is 0.30~dex and it is fully supported by 
the intrinsic errors of the different samples (see the error bars
in the top right corner).

We also found that a quadratic relation allows us to calibrate
the SDSS-SSPP iron abundances into the HR metallicity scale:  
\begin{eqnarray}
{\tt FEHADOP}^*=-0.65+0.60\cdot({\tt FEHADOP}+0.26) \\
-0.05\cdot ({\tt FEHADOP}+0.26)^2 \nonumber
\end{eqnarray}
Once this relation was applied to the SSPP iron abundances we 
obtained a null residual with a dispersion of 0.27~dex.

%_______________________________________________________________________________
\subsection{{Validation of the spectroscopic measurements: individual vs co-added spectra}}

We have already mentioned in Sect.\ref{sec:segueData}, that the metallicity
estimates rely on the application of the \deltaS\ method to the SDSS-SEGUE
co-added spectra. The co-added spectrum is typically based on three back-to-back
900s individual spectra collected but, in order to achieve highest
signal-to-noise ratio, the individual spectra can also spread over days
\citep{bickerton12}. Moreover, it is worth mentioning that the spectra were
collected at random pulsation phases. To quantify the impact that the co-adding
of spectra collected at random pulsation phases have on the metallicity
estimates, we evaluated the difference between the metallicity estimates based
on the co-added spectrum with the metallicity based on the application of the
\deltaS\ method on the individual spectra. To validate the approach, we selected
the individual spectra with a signal-to-noise ratio larger than $\sim$20. We
ended up with a sample of more than 1,000 RRab variables.

\begin{figure}[b]%%%%%%%
\centering
\includegraphics[width=\columnwidth]{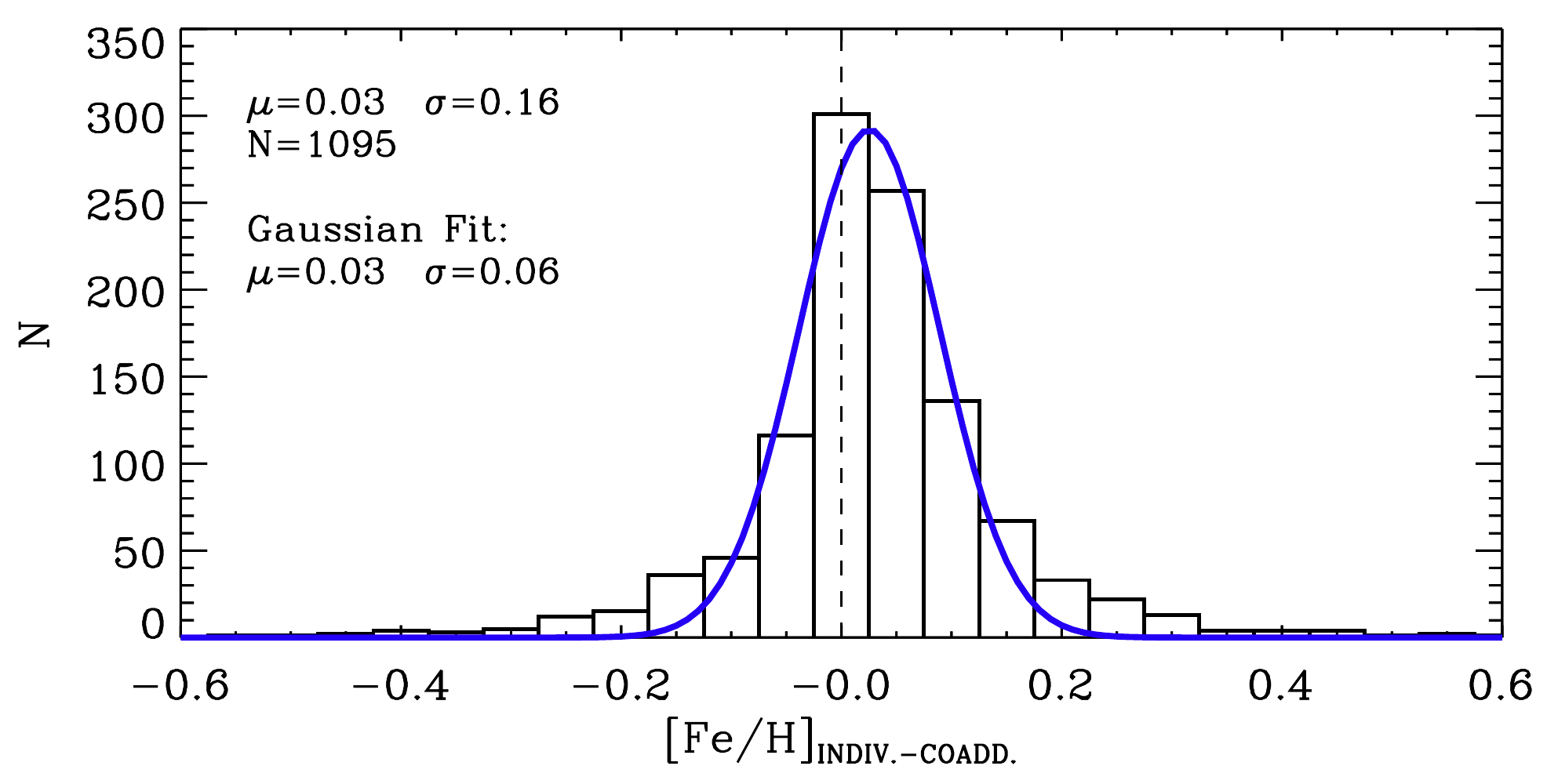} 
\caption{Distribution of the difference between the mean of the individual 
metallicity measurements and the metallicity measured on the co-added
spectrum for a sample of 1,095 RRLs. The blue curve shows the Gaussian fit
to the distribution. The mean and the standard deviation of the sample
and of the Gaussian fit are also labelled.}\label{fig:deltas_repeated}
\end{figure} %%%%%%%

Figure~\ref{fig:deltas_repeated} shows the distribution of the difference
between the mean of individual \feh\ estimates and the \feh\ measured on the
co-added spectrum. The Gaussian fit to the distribution (blue curve) gives a
peak of $\sigma$=0.06~dex, while the standard deviation of the measurements is
0.16~dex, i.e. a factor of two smaller than the standard deviation of the
calibration of the \deltaS\ method with the HR metallicity scale (0.29~dex, see
also Sect.\ref{sec:catalogue}). Moreover and even more importantly, the current
evaluation agrees quite well with similar similar estimated provided by
\citet[][0.22~dex]{drake13a} by using SDSS spectra. In passing we note that the
modest value in the mean difference further supports the use the co-added
spectra, typically characterised by higher signal-to-noise ratios, to determine
metal abundances of RR Lyrae by using the \deltaS\ method.
%_______________________________________________________________________________
\subsection{Validation of the spectroscopic measurements: V~Ind}

It is worth noting that the SDSS-SEGUE spectroscopic data were collected at
random phases along the pulsation cycle, with exposure times of 15~min
\citep{smee13}. This means that a fraction of the spectra could have been
collected along the rising branch. The rising branch has always been avoided in
the spectroscopic analysis and in the application of the Baade-Wesselink method
\citep{storm94}. The reasons are manifold. Dating back to more than half century
ago, \citet{preston64} demonstrated on empirical basis that across these phases
a strong shock is formed and propagates towards the outermost regions. This
causes the occurrence of line doubling and P~Cygni profile, further
supporting the presence of strong nonlinear phenomena in the outermost layers.
This empirical scenario was soundly supported by nonlinear, convective models
taking account for time dependent convective transport suggesting that the
efficiency of the convective transport attains its maximum efficiency along the
rising branch. This is not a severe limitation, since the time interval between
minimum and maximum light is of the order of 10\%\ of the pulsation cycle.
However, these are the reasons why the \deltaS\ method was not applied to this
portion of the pulsation cycle \citep{freeman75,layden93}. In dealing with large
spectroscopic samples we cannot exclude that a minor fraction can also be
collected during these pulsation phases. Moreover, we still lack quantitative
constraints of the impact that these phenomena have on abundance estimates based
on the \deltaS\ method.  

\begin{figure} %%%%%%%
\centering
\includegraphics[width=\columnwidth]{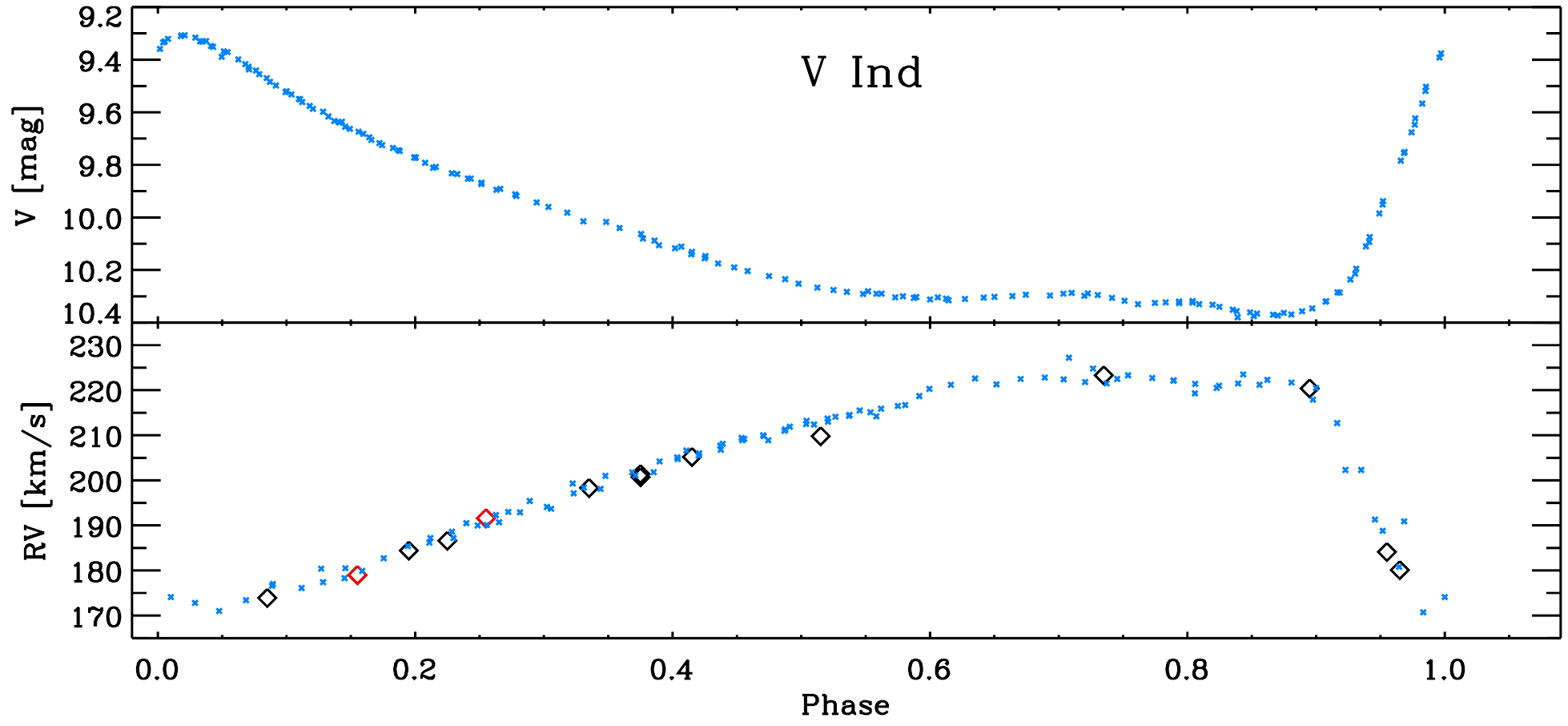} 
\caption{Visual light curve of V~Ind (top panel) and radial velocity curve
(bottom panel) as function of the pulsation phase \citep[blue crosses][]{clementini90}. 
Black diamonds mark the radial
velocities based on X-shooter spectra \citep{magurno18phd}, while the red ones
are used for UVES ones \citep{pancino15}.}\label{fig:vind1}
\end{figure} %%%%%%%

\begin{figure} %%%%%%%
\centering
\includegraphics[width=\columnwidth]{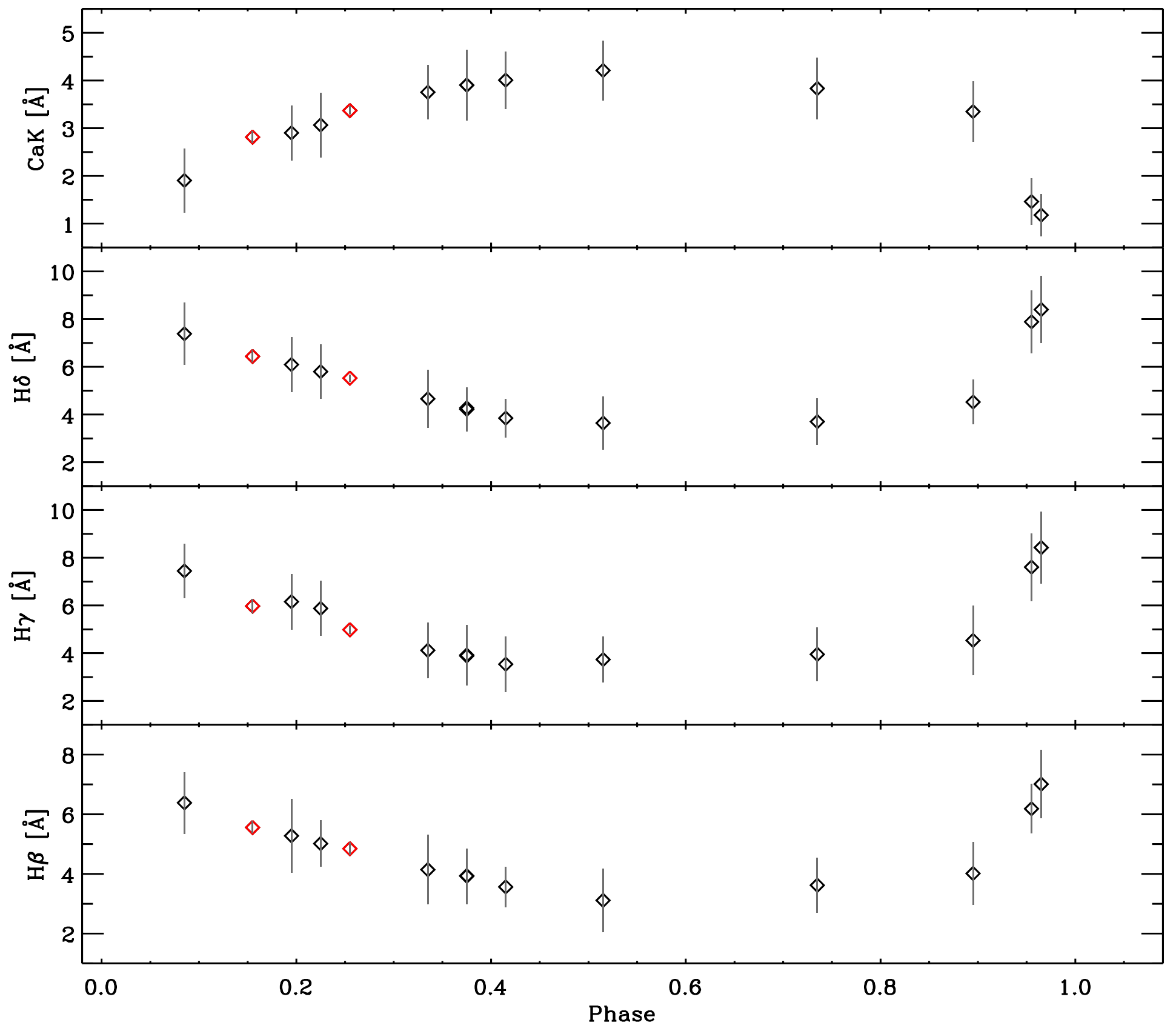} 
\caption{Equivalent widths of the four spectroscopic diagnostics 
adopted to apply the \deltaS\ method to fundamental RRL V~Ind 
as a function of the pulsation phase. The symbols are the same 
as in Fig.~\ref{fig:vind1}.}\label{fig:vind2}
\end{figure} %%%%%%%

\begin{figure} %%%%%%%
\centering
\includegraphics[width=\columnwidth]{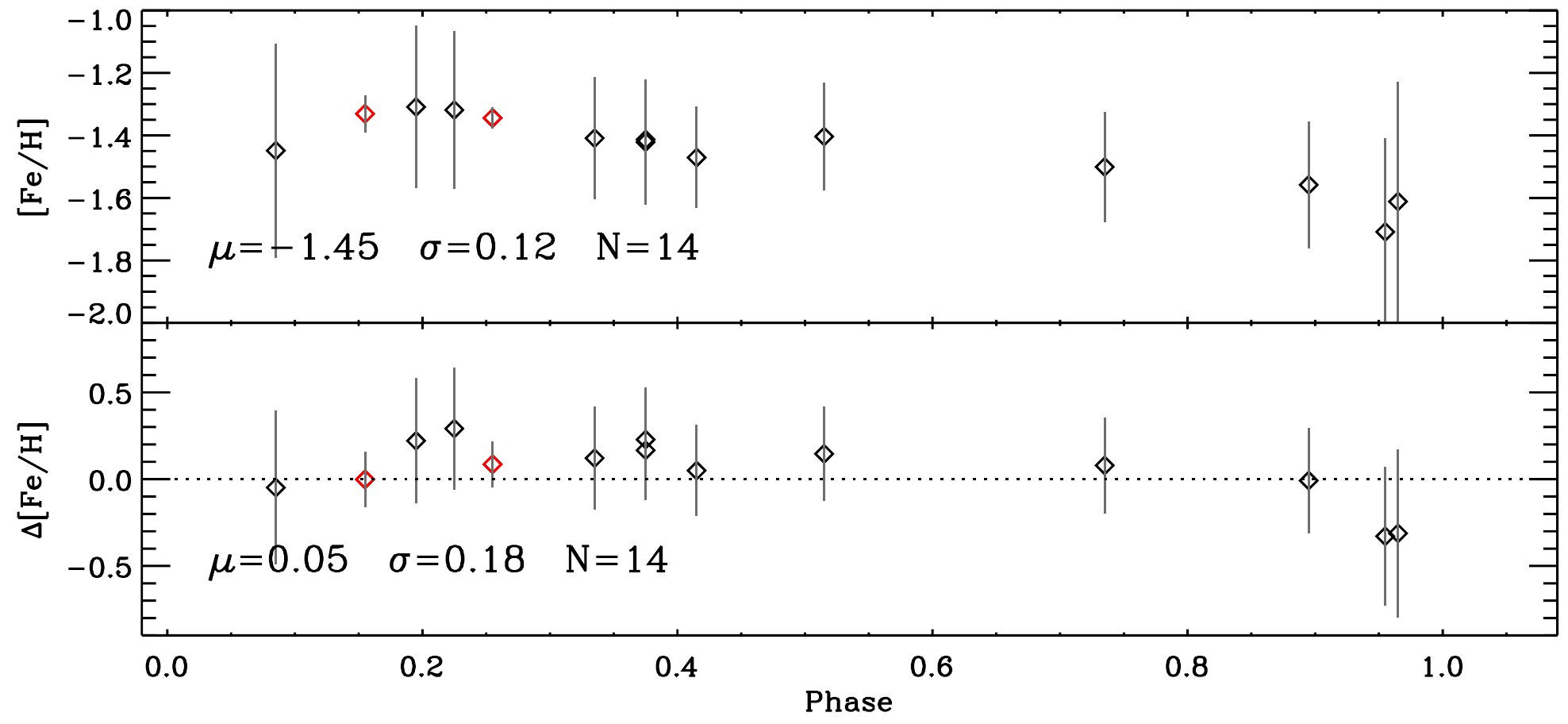} 
\caption{\textit{Top:} iron abundances for V~Ind based on the 
\deltaS\ method. \textit{Bottom:} difference in iron abundance with 
the iron values provided by \citet{magurno18phd} and by 
\citet{pancino15}. The symbols are the same 
as in Fig.~\ref{fig:vind1}.}\label{fig:vind3}
\end{figure} %%%%%%%

In a recent investigation, \citet{magurno18phd} estimated iron abundance of a
field, short-period ($P$$\sim$0.48~day), large-amplitude (${\rm A}_V$$\sim$1.07~mag,
\citealt{monson17}) fundamental RRL: V~Ind. He adopted twelve medium-resolution
(R$\sim$10,000-18,000), high signal-to-noise ratio ($\sim$200) spectra
collected with X-shooter \citep{vernet11} at ESO/VLT\footnote{Based on
observations collected under ESO programme ID 297.D-5047(A), PI. G. Bono.}. The
key advantage of X-shooter is the possibility to simultaneously cover a very
wide wavelength regime, ranging from $\sim$3000 to $\sim$25,000\AA. Moreover,
the spectra cover the entire pulsation cycle and the exposure times are quite
short (120-180 sec). The target is quite bright (\vv=9.97~mag) for an 8m class telescope and we
decided to use a narrow slit of 0.4\arcsec, obtaining a spectral resolution in
the optical range of the order of 18,000. It is worth mentioning that V~Ind is an acid 
test to investigate the metallicity estimates along the pulsation cycle, 
since it is among the RRLs with the largest pulsation amplitudes. On the basis of 
these spectra,
\citet{magurno18phd} measured the iron abundance of V~Ind as function of its
pulsation phase. Fig.~\ref{fig:vind1} shows the visual light curve (top) and
the radial velocity (bottom) as function of the pulsation phase
\citep[blue crosses from][]{clementini90}. The radial velocity measurements based on X-shooter
spectra are marked with black diamonds. To improve the sampling along the
pulsation cycle, we also included two high-resolution (R$\sim$40,000) spectra
collected with UVES \citep{dekker00} at ESO/VLT\footnote{Based on observations
collected under ESO programme ID 083.B-0281(A), PI. D. Romano.} and available in
the ESO science archive (red diamonds in Fig.~\ref{fig:vind1}). 
Fortunately enough, both the X-shooter and the UVES spectra cover the wavelength
range from \caiik\ to \hbeta\ lines.

To validate the adopted \deltaS\ method as function of the pulsation phase, the
quoted high-resolution spectra were degraded and re-binned to the spectral
resolution (R$\sim$2000) and sampling ($\Delta\log{\lambda}$=0.0001) of
SDSS-SEGUE spectra. The quality of the re-binned spectra was quite good, with a
signal-to-noise ratio of $\sim$200 and they appear to be quite similar to the
best SEGUE spectra (see Fig.~\ref{fig:spectra}). We applied the \deltaS\ method
described in Sect.~\ref{sec:deltas} to the re-binned spectra and the same
spectroscopic calibration described in Sect.~\ref{sec:calib} to transform the EWs
into iron abundance. Fig.~\ref{fig:vind2} shows the EW measurements of the
spectral features involved in the \deltaS\ method and they show the expected
trend in \caiik\ and H lines. Data plotted in the top panel of this figure
clearly show that the EW of the \caiik\ line steadily increases when moving
along the decreasing branch, it attains its maximum across the phases of minimum
light and it starts to decrease along the rising branch. 
The trend for the H lines is exactly a mirror image of
the \caiik\ line. This means that the ratio between the EWs of \caiik\ and H
lines remains almost constant over the entire cycle. This is the reason why the
\feh\ abundances based on \deltaS\ method do not show a phase dependence.
Indeed, the iron abundances attain similar values, within the errors, over the
entire pulsation cycle. 

The top panel of Fig.~\ref{fig:vind3} shows the \feh\ estimates as a function
of pulsation period: the mean value is $-1.45$~dex, while the standard deviation
of the measurements is 0.12~dex. The current mean iron abundances agree quite
well similar estimates provided by \citet{magurno18phd} and by \citet{pancino15}. In
fact, the bottom panel of Fig.~\ref{fig:vind3} shows the phase-to-phase
difference in iron abundance based on \deltaS\ compared to those from
\citet{magurno18phd} and \citet{pancino15} for X-shooter and UVES data,
respectively. The comparison indicates that iron abundances based on \deltaS\
method and those based on high-resolution spectra attain similar values, indeed
the mean is vanishing (0.05~dex), while the standard deviation is 0.18~dex. 
More important, the \feh\ values agree within the errors also along the raising 
branch of V~Ind. Finally, the quoted results allow us to use the derived \feh\ 
abundances from the \deltaS\ method, independently of the pulsation phase.

\begin{table*}%%%%%%%%%%%%%%%%%%%%%%%%%%%%%%%%%%%
\caption{Number of objects per dataset included in the iron catalogue.\label{tab:catalogue}}
\scriptsize
\begin{center}
\begin{tabular}{lcccccccc|r}
          &\textbf{SEGUE}&\textbf{Magurno+18}&\textbf{NGC~5272}&\textbf{Rave DR5}&\textbf{Sesar+13}& \textbf{Dambis+13}&\textbf{Duffau+14}&\textbf{SSPP} & \textbf{$\epsilon$\feh}\\
          \tableline
\textbf{SEGUE}&        2382&           0&           5&           0&          21&           6&          18&           0&0.29\\
\textbf{Magurno+18}&\dotfill&         104&           0&           0&           0&          72&           0&           0&0.10\\
\textbf{NGC~5272}&\dotfill&\dotfill&           5&           0&           0&           0&           0&           0&0.15\\
\textbf{Rave DR5}&\dotfill&\dotfill&\dotfill&           6&           0&           2&           0&           0&0.20\\
\textbf{Sesar+13}&\dotfill&\dotfill&\dotfill&\dotfill&          50&           0&           0&           1&0.15\\
\textbf{Dambis+13}&\dotfill&\dotfill&\dotfill&\dotfill&\dotfill&         360&           0&           0&0.22\\
\textbf{Duffau+14}&\dotfill&\dotfill&\dotfill&\dotfill&\dotfill&\dotfill&          57&           1&0.15\\
\textbf{SSPP}&\dotfill&\dotfill&\dotfill&\dotfill&\dotfill&\dotfill&\dotfill&          65&{\tt FEHADOPUNC\tablenotemark{a}}\\
	\tableline
\end{tabular}
\end{center}
\tablenotetext{a}{The error was summed in quadrature with the dispersion 
of the residuals obtained from the comparison with the calibrating sample 
(0.27~dex).}
\end{table*}%%%%%%%%%%%%%%%%%%%%%%%%%%%%%%%%%%%

%_______________________________________________________________________________
\section{Metallicity distribution}\label{sec:catalogue}

We already mentioned that we are dealing with a sample of 2,903 RRLs on the same
metallicity scale. Note that for objects that belong to different data sets we
are adopting the following priority. The iron abundances based on the \citet{magurno18} 
sample (104) were included with their original estimates and intrinsic errors. For the
RRLs for which the error was not provided, we assumed a mean error of 0.1~dex.
The original iron abundances were also included for the five cluster RRLs and the
six RRLs retrieved from the Rave DR5 catalogue. The former sample has an intrinsic
error of 0.15~dex, while the latter one has an intrinsic error of 0.20~dex. The
RRLs for which the iron abundance is based on the current \deltaS\ method (2,382)
come immediately after in the priority list, and the error for this sample was
assumed equal to the standard deviation of the calibration with the HR
metallicity scale (0.29~dex). This is the largest and most homogenous sample of
RRL iron abundances ever estimated. These two samples were complemented with RRL
iron abundances provided by \citet[][50 stars]{sesar13}, by \citet[][360
stars]{dambis13} and by \citet[][57 stars]{duffau14}. 
For these samples the error on individual measurements was
estimated by assuming a mean error of 0.15, of 0.22 and 0.15~dex, respectively.
Finally, we added the RRL iron abundances provided by the SDSS-SSPP survey
(65). The error on individual measurements were estimated by summing in
quadrature the original uncertainties \texttt{FEHADOPUNC} with the standard
deviation of the calibration with the HR metallicity scale (0.27~dex, see
Sect.~\ref{sec:calib}). The final iron abundance of RRLs in common among
different medium/low resolution data sets (\deltaS, \citealt{sesar13},
\citealt{dambis13}, \citealt{duffau14} and SSPP) was estimated as the mean of the different
measurements and the errors were summed in quadrature. In
Table~\ref{tab:catalogue} are listed the number of stars in common between the
different datasets.

\begin{figure} %%%%%%%
\centering
\includegraphics[width=\columnwidth]{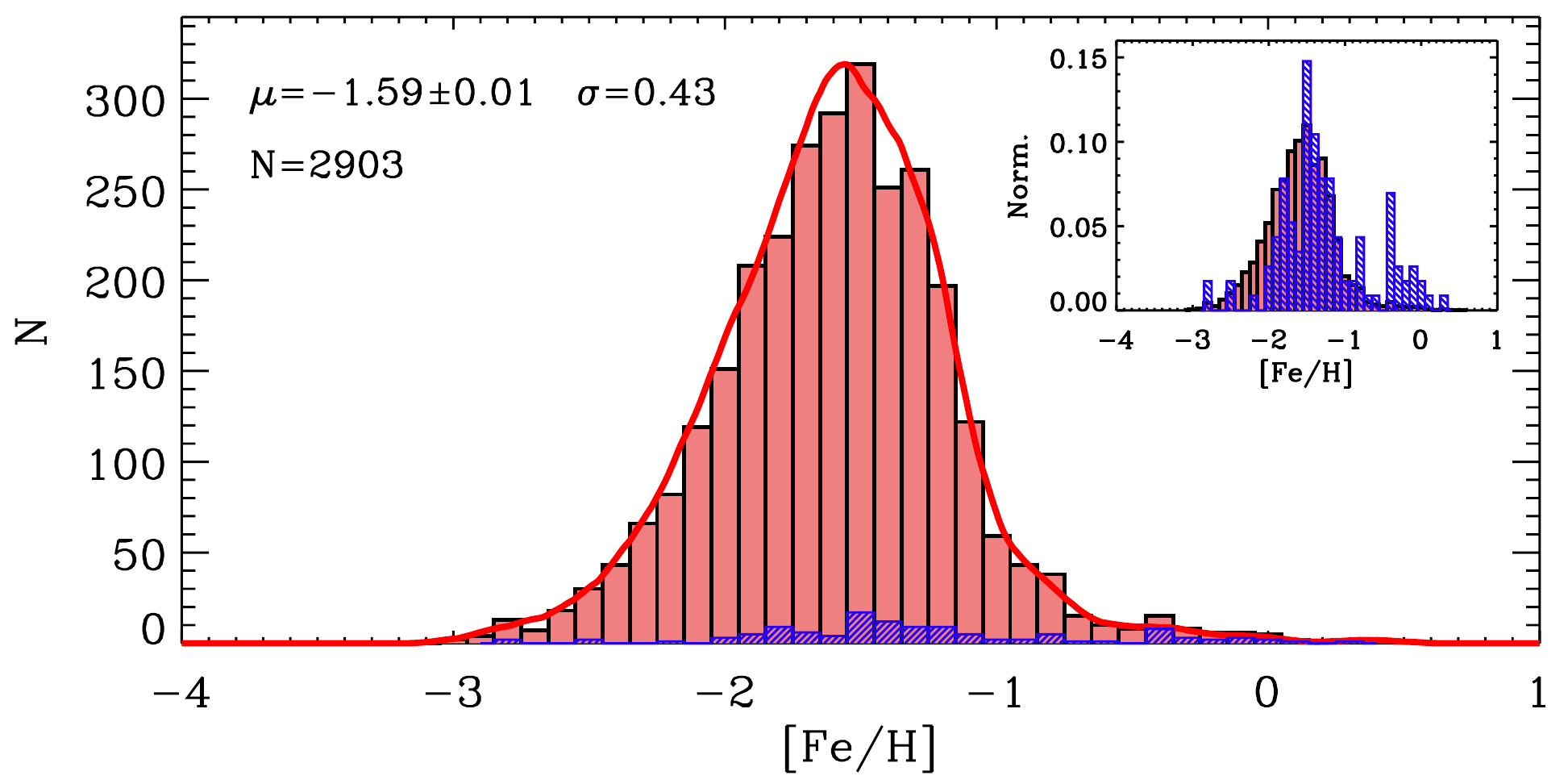} 
\caption{Metallicity distribution of the entire spectroscopic sample (in red) 
and for the high resolution sample (blue histogram).
The orange line shows the smoothed metallicity distribution. The 
inset shows the same metallicity distributions, but area 
normalised.}\label{fig:distr_feh}
\end{figure} %%%%%%%

Fig.~\ref{fig:distr_feh} shows the metallicity distribution of the entire RRab
spectroscopic sample (red histogram) together with the smoothed metallicity
distribution (orange line). The latter one was smoothed using a Gaussian kernel
with unitary weight and $\sigma$ equal to the error of the individual estimates.
The mean and the standard deviation of the smoothed distribution are also
labelled. Data plotted in this figure bring forward several interesting features
worth being discussed in detail. 

{\em a) Mean and Standard Deviation} -- The spectroscopic sample we are dealing
with is more than a factor of five larger than any previous spectroscopic
investigation of field RRLs
\citep{dambis13,kinman12,layden93,layden94,layden95}. The current mean metal
abundance agrees quite well, within the errors, with similar estimates available
in the literature (\feh=--1.59 \textit{vs} --1.65, \citealt{layden94}). 
The same outcome applies to the standard deviation, indeed, 
the difference ($\sigma$=0.43 \textit{vs} 0.34~dex) is once
again marginal if we take account for the difference in the sample size.   

{\em b) Tails} -- The metallicity distribution appears more skewed toward the
metal-poor regime, indeed, the metal-poor tail approaches \feh$\simeq$--3, while
the metal-rich one approaches Solar iron abundance. The quoted metallicity range
is also supported by iron abundances based on high-resolution (HR) spectra (blue
dashed areas). The main difference between the HR abundances and those based on
lower-resolution (LR) spectra is that the former ones show a more prominent
metal-rich tail and a less prominent metal-poor tail when compared with the
latter ones. A glance at the metallicity distributions plotted in the inset 
of the same figure, that are normalised according to the area, shows even more 
clearly the difference in the metal-poor/metal-rich tails. On the basis of the 
current data it is not clear whether the metal-poor tail based on LR spectra 
might be a drift of the current absolute calibration or intrinsic. Note that 
the metallicity regime more metal-poor than \feh$\sim$--2.3 is not covered 
by cluster RRLs and the number of field, very metal-poor RRLs for which iron 
abundances is based on HR spectra is still limited (four). This limitation 
applies if we also take account for RRc variables. 

{\em c) Magnitude distribution} -- The HR sample is only limited to bright
nearby RRLs, while the whole sample covers more than 120~kpc (see
Fig.~\ref{fig:distr_magn}). This indicates that the difference in the
metallicity distribution between iron abundances based on either HR spectra 
or \deltaS\ method might also be caused by an observational bias affecting
the former sample.

%_____________________________________________________________________
\section{The fine structure of the Bailey~diagram}
\label{sec:per-met}

The reasons why the Bailey diagram is a useful diagnostic to investigate the
pulsation properties of variable stars have already been mentioned in the
Sect.~\ref{sec:intro}. Here we only mention two relevant key points: \textit{a)}
it is independent of uncertainties affecting distance and reddening; \textit{b)}
cluster RRLs can be split
into two groups called "Oosterhoff I" (OoI, mean
RRab period of 0.55 days) and "Oosterhoff II" (OoII. mean RRab period of 0.65
days). The pioneering investigations concerning the metal content of globular
clusters by \citet{arp55} and by \citet{kinman59} clearly demonstrated that OoI
globulars are more metal-rich than OoII globulars. A quantitative investigation
of the dependence of the Oosterhoff dichotomy on the metal content has been
hampered by two intrinsic properties of Galactic globulars: 

{\em a)} The metallicity distribution of Galactic globulars is bimodal
\citep{harris91} with a well defined minimum for \feh=--0.8/--1.0. Moreover, the
metal-poor tail does not approach the limit of field Halo stars, while the
metal-rich tail does not approach the limit of old metal-rich Bulge stars. The
difference is well known and it is tightly connected with the formation
mechanism of globular clusters \citep{choksi18}.

{\em b)} Galactic globulars display at fixed metal content relevant changes in
the horizontal-branch (HB) morphology, the so-called "second parameter" problem.
This means that at fixed metal-content they might or they might not host RRLs
according to their HB morphology. Nearby dwarf galaxies do not help in 
unraveling the skein, because their HB morphologies are quite similar. Indeed, 
only a few of them host an old stellar component that is either more metal-rich than 
\feh=--1.0 or more metal-poor than \feh=--2.2 (see Fig.~12 in \citealt{mcconnachie12}).
Quite often these stellar systems have been classified as Oosterhoff
intermediate, i.e. the RRab attain mean periods that are between OoI and OoII
clusters. The reader interested in detailed discussion concerning the difference
among different globulars and nearby dwarf galaxies is referred to
\citet{fiorentino17} and \citet{braga18}, and references therein. 

\begin{figure} %%%%%%%
\centering
\includegraphics[width=\columnwidth]{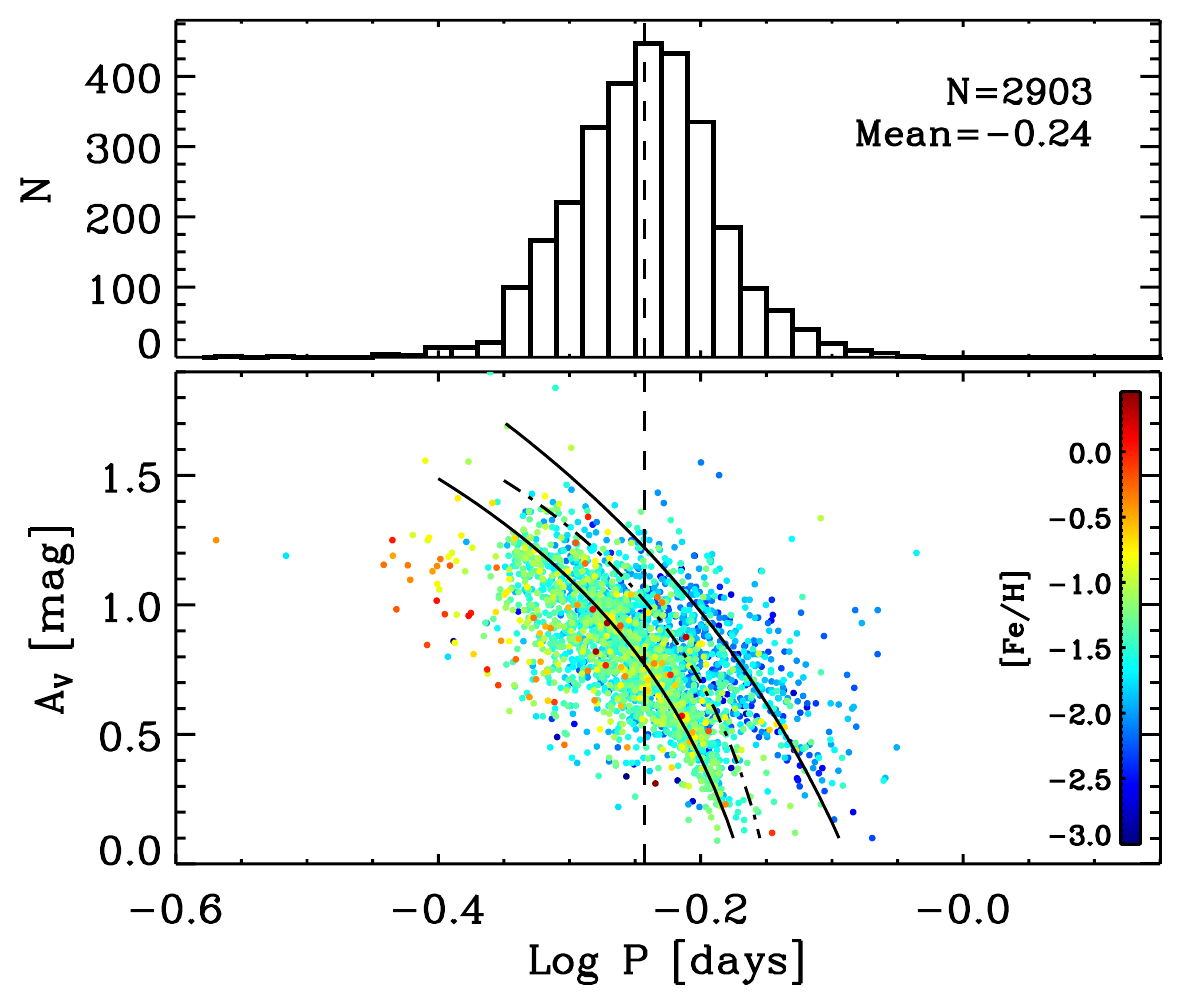} 
\caption{\textit{Top:} Period distribution of the entire spectroscopic sample.
\textit{Bottom:} Bailey diagram of the spectroscopic sample. The metallicity 
is colour coded and the colour bar is plotted on the right. 
The vertical dashed line marks the mean period of the entire sample. 
The solid lines display the new analytical relations for OoI and OoII 
overdensities. The dot-dashed line shows the Oosterhoff intermediate loci, 
defined as the "valley" between the two main overdensities.}\label{fig:bailey}
\end{figure} %%%%%%%

The quoted circumstantial evidence indicates that we still lack a homogenous
and detailed analysis of the Bailey diagram as a function of the metal content.
In this context it is worth mentioning that we are neglecting the metallicity 
estimates based either on photometric indices such as the inversion of the 
PL relation \citep{braga16,martinez16,bono19} or the Fourier decomposition of 
the light curve \citep{jurcsik96,nemec13,elorrieta16,hajdu18}. 
Data plotted in Fig.~\ref{fig:bailey} open a new path concerning the 
dependence of the luminosity amplitude on metallicity. 

\begin{figure*} %%%%%%%
\centering
\includegraphics[width=1.8\columnwidth]{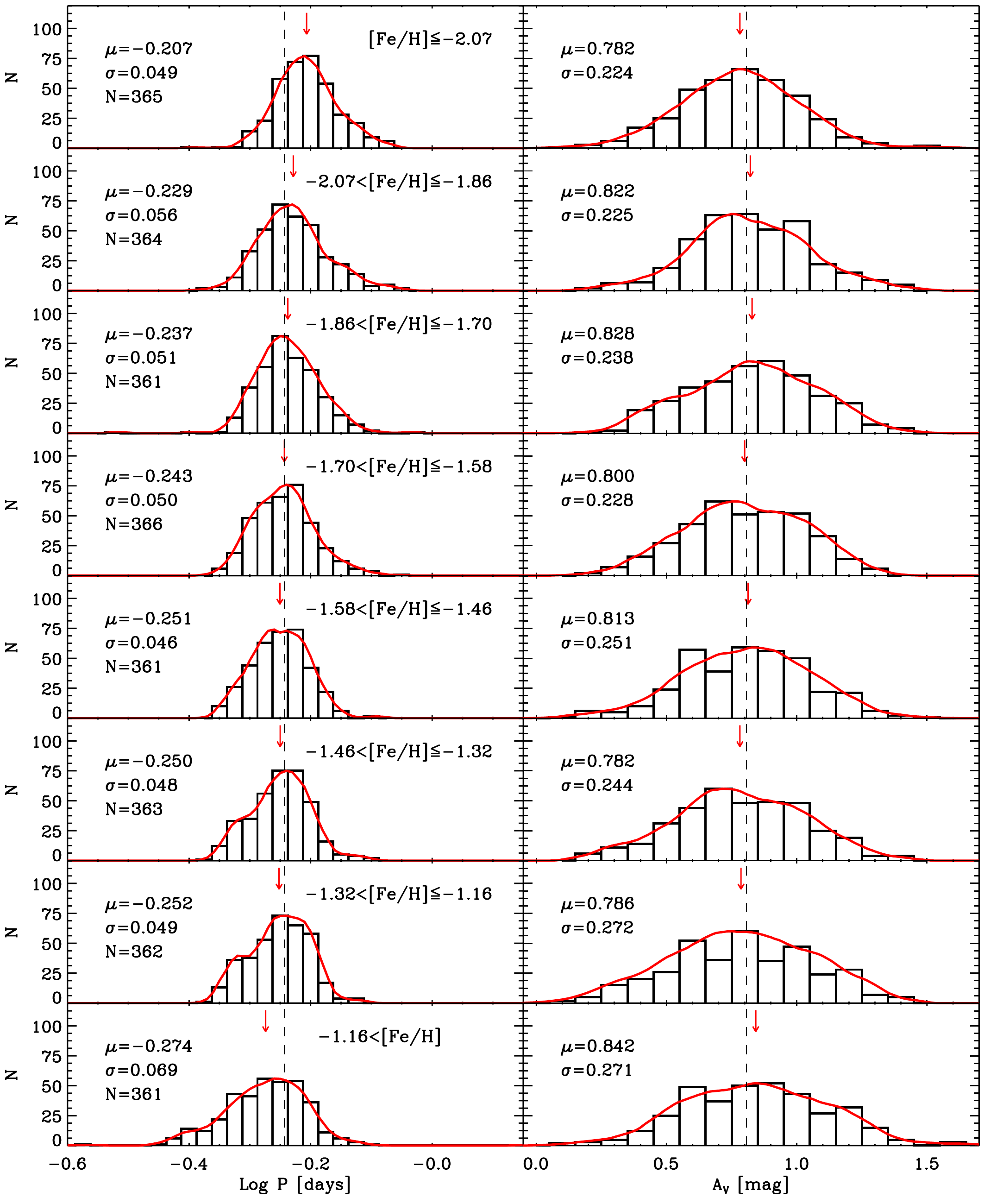} 
\caption{Period (left panels) and visual amplitude (right panels) 
distributions of the spectroscopic sample. The red lines display the 
smoothed distributions. The sample was split in eight 
metallicity bins including a similar number of objects (see labelled values). 
The red arrows mark the mean period and the mean amplitude of the 
individual bins. The dashed lines show the mean $\log P$ and mean 
$A_V$ of the total sample.}\label{fig:perfeh}
\end{figure*} %%%%%%%

{\em a) Period-metallicity correlation} -- The metallicity is colour coded (see
the bar on the right axis) and indicates that RRab variables become, on average,
steadily more metal-rich when moving, at fixed amplitude, from the long- to the short-period regime.
The trend was already known, but the current data are suggesting that the 
variation of the mean period of RRab variables is continuous, i.e. the 
distribution of the RRab variables in the Bailey diagram is not uniform, 
but the variation is far from being dichotomic.
To trace the key features of the Bailey diagram we produced a 3D histogram
($A_V$, $\log$P, number of RRLs) with the entire RRL sample. We traced the local
maxima and the local minima in this 3D diagram and then we smoothed them by
applying a running average. The two solid lines display the "mean" locus of the
local maxima associated to OoI and OoII clusters, while the dashed line traces
the Oosterhoff intermediate loci, defined as the local minima between the two
main overdensities.
The analytical relations for the three Oosterhoff sequences are the following:  
\begin{eqnarray}
{\rm OoI:} &A_V&= 2.62+2.08\cdot\log(-0.11-\log P) \\
{\rm OoII:} &A_V&= 3.13+3.48\cdot\log(0.041-\log P)\\
{\rm OoInt:} &A_V&= 2.57+1.72\cdot\log(-0.12-\log P)
\end{eqnarray}

The quoted relations are in good agreement with similar
relations for OoI and OoII groups provided by 
\citet{zorotovic10} and based on cluster RRLs collected 
by \citet{cacciari05}. The mean difference in luminosity 
amplitude, over the entire period range, is $\sim$0.2~mag 
for the OoI group and $\sim$0.1~mag for OoII group.

Note that the Oo sequences are far from being parallel when moving from the
short- to the long-period regime. Moreover, the region of the Bailey diagram
among the two Oo sequences with periods of the order of $\log P$$\sim$--0.15 and
amplitudes smaller than 0.5~mag appears empty. This suggests that this region 
is a sort of avoidance region for a broad range of metal abundances.   

{\em b) Period-amplitude-metallicity correlation} -- The RRab cover, at fixed
metallicity and luminosity amplitude, a broad period range. This means that
period, amplitude and metallicity do not obey to simple linear correlations.    

To constrain on a more quantitative basis the variation of the pulsation
properties (period, luminosity amplitude) as a function of the metal content we
divided the entire spectroscopic sample in eight different metallicity bins. The
edges of the individual metallicity bins (see labeled values in
Fig.~\ref{fig:perfeh}) were changed in such a way that they include a similar
number of RRab variables. The left panels of Fig.~\ref{fig:perfeh} display from
top to bottom the period distribution of RRab variables from the metal-poor to
the metal-rich tail. The mean period, the standard deviation and the number of
RRLs per metallicity bin are also labeled. The right panels show the \vv\
amplitude distributions of the same RRLs plotted in the left panels. The period
and the amplitude distributions display several interesting features:  

{\em a)} The mean period becomes systematically shorter when
moving from the metal-poor to the metal rich regime. Indeed, the red arrow 
moves from the right to the left of the mean period of the entire sample 
(vertical dashed line). 

{\em b)} The period distribution is asymmetric over the entire metallicity
range, but the skewness of the distribution moves from the long to the short
period range when moving from the metal-poor to the metal-rich regime. The
standard deviation of the different period distribution is quite constant, 
but the period distribution in the metal-rich regime becomes steadily 
flatter.   

{\em c)} The luminosity amplitudes do not display the linear trend 
found for the pulsation periods. Indeed, the mean luminosity 
amplitude shows a modest variation and it moves either to slightly 
smaller or to slightly larger values of the mean global amplitude in 
the different metallicity bins. This evidence is suggesting that the
dependence of the luminosity amplitude appears to be significantly 
milder than the dependence of the pulsation period. In passing, 
we also note that the large amplitude tail becomes, as expected, 
more and more relevant in the metal-rich regime (HASPs).   

\begin{figure} %%%%%%%
\centering
\includegraphics[width=\columnwidth]{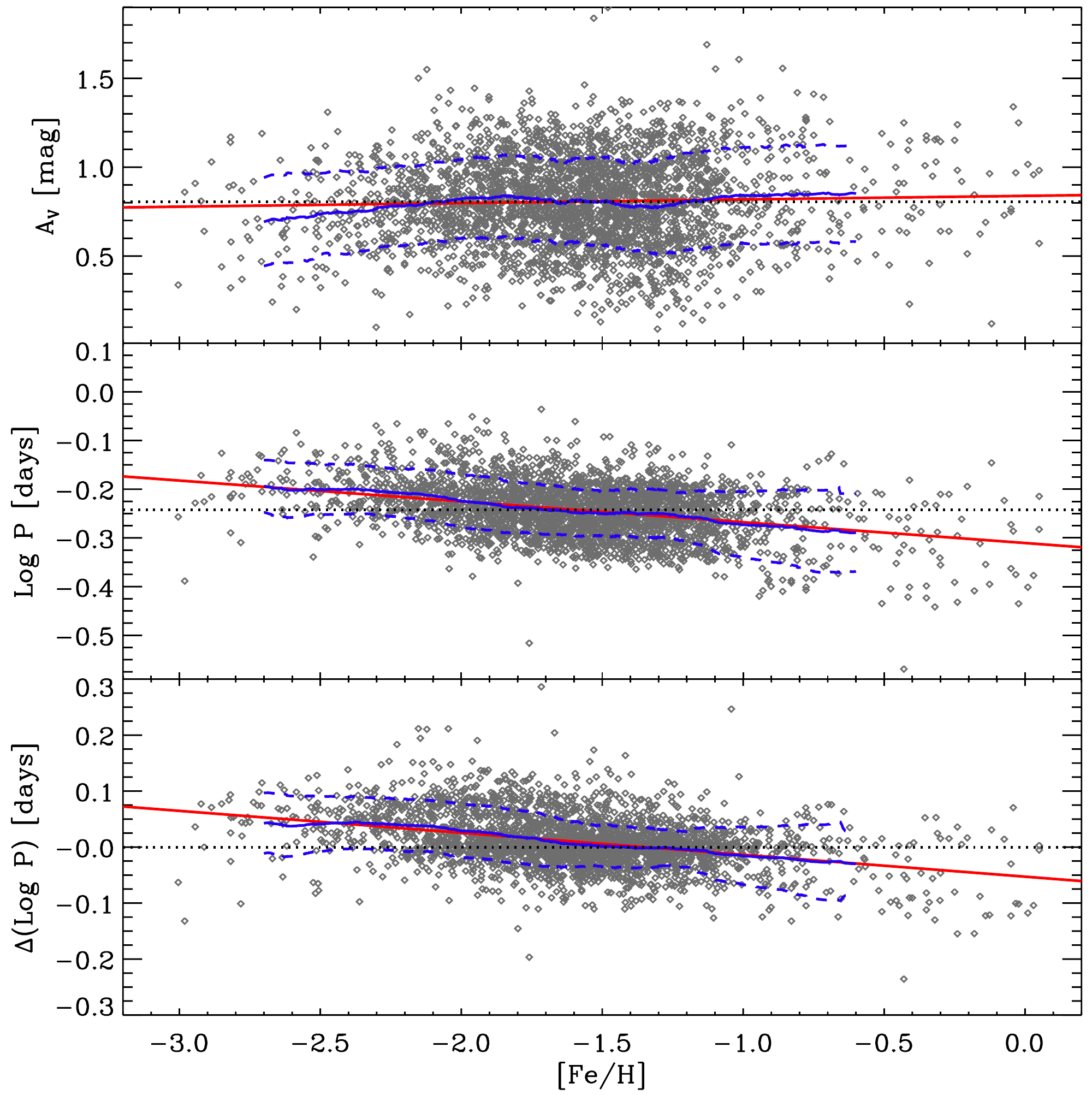} 
\caption{\textit{Top:} \vv\ amplitude as function of \feh. A running average
(blue) and a linear regression (red) are also displayed. \textit{Middle:} The
same as the top but with the $\log P$ on y-axis. \textit{Bottom:} The same as
the top but with the $\Delta\log P$ on y-axis, i.e. the difference in period
with the OoI relation. 
The horizontal dotted lines mark the mean values of {\it y}-axis}.\label{fig:perfehfit}
\end{figure}

To overcome the limitation in the number of metallicity bins and possible 
subtle fluctuations in correlation between the two pulsation parameters 
and the metallicity we performed a running average. The entire sample of 
RRab variables was ranked as a function of the metal content and we 
estimated the running average, with a running box containing 500 objects.
Note that in this estimate we neglected the very metal-poor (\feh$\le$--2.7) and
the very metal-rich (\feh$\ge$--0.4) tail, due to the poor statistics in these
metallicity ranges. The metallicity and the mean visual amplitude of the bin
were estimated as the mean over the individual iron abundances and visual
amplitudes of the 500 objects included in the box. We estimated the same
quantities moving by one object in the ranked list until we took account for the
last object in the sample with the most metal-rich abundance. The solid
blue line plotted in the top panel of Fig.~\ref{fig:perfehfit} shows the
running average, while the two dashed lines display the 1$\sigma$ standard
deviation. 

 %%%%%%%

The linear fit (red line) plotted in the same panel shows a mild increase in the
visual amplitude when moving from the metal-poor to the metal-rich regime. The
linear relation fitting the data is the following:
\begin{equation}
A_V=0.84(\pm 0.02)+0.02(\pm 0.01) \cdot \feh
\end{equation}

However, the difference with the mean global amplitude (horizontal dotted line)
is of the order of a few hundredths of a magnitude. Indeed, the current fit 
suggests that a variation of $\approx$2~dex in metallicity causes a
variation of $\sim$0.04 magnitudes in visual amplitude. The current findings
clearly indicate that the association of a luminosity amplitude to an iron
abundance should be cautiously treated, indeed, at fixed visual amplitude, field
RRLs cover more than 2~dex in metal content.  

The dependence of the mean period on the iron abundance is more solid, and
indeed, the mean period decreases from 0.63 days in the metal-poor regime
(\feh$\sim$--2.5) to 0.51 days in the metal-rich (\feh$\sim$--0.5). This means a
steady decrease of 0.12~days over a variation of 2~dex in metallicity. 
We evaluated the running average values (blue solid line in the middle panel 
of Fig.~\ref{fig:perfehfit}) and performed a linear fit (red line) finding: 
\begin{equation}
\log P = -0.311(\pm 0.004) - 0.044(\pm 0.002) \cdot \feh
\end{equation}
The linear variation of the mean period as a function of the metallicity and
the similarity of the standard deviation over the entire sample is further
supporting the smooth variation of this intrinsic parameter when moving from the
metal-poor to the metal-rich regime. To constrain on a more quantitive basis
possible variations among metal-poor, metal-intermediate and metal-rich regime,
we also estimated the difference in period between individual RRLs and the Oo~I
analytical relation derived in Sect.~\ref{sec:per-met}. Data plotted in the
bottom panel of Fig.~\ref{fig:perfehfit} show, once again, a smooth variation
over the entire metallicity range, with the following linear relation:
\begin{equation}
\Delta \log P = -0.054(\pm 0.003) - 0.040(\pm 0.002) \cdot \feh
\end{equation}

\begin{figure} %%%%%%%
\centering
\includegraphics[width=\columnwidth]{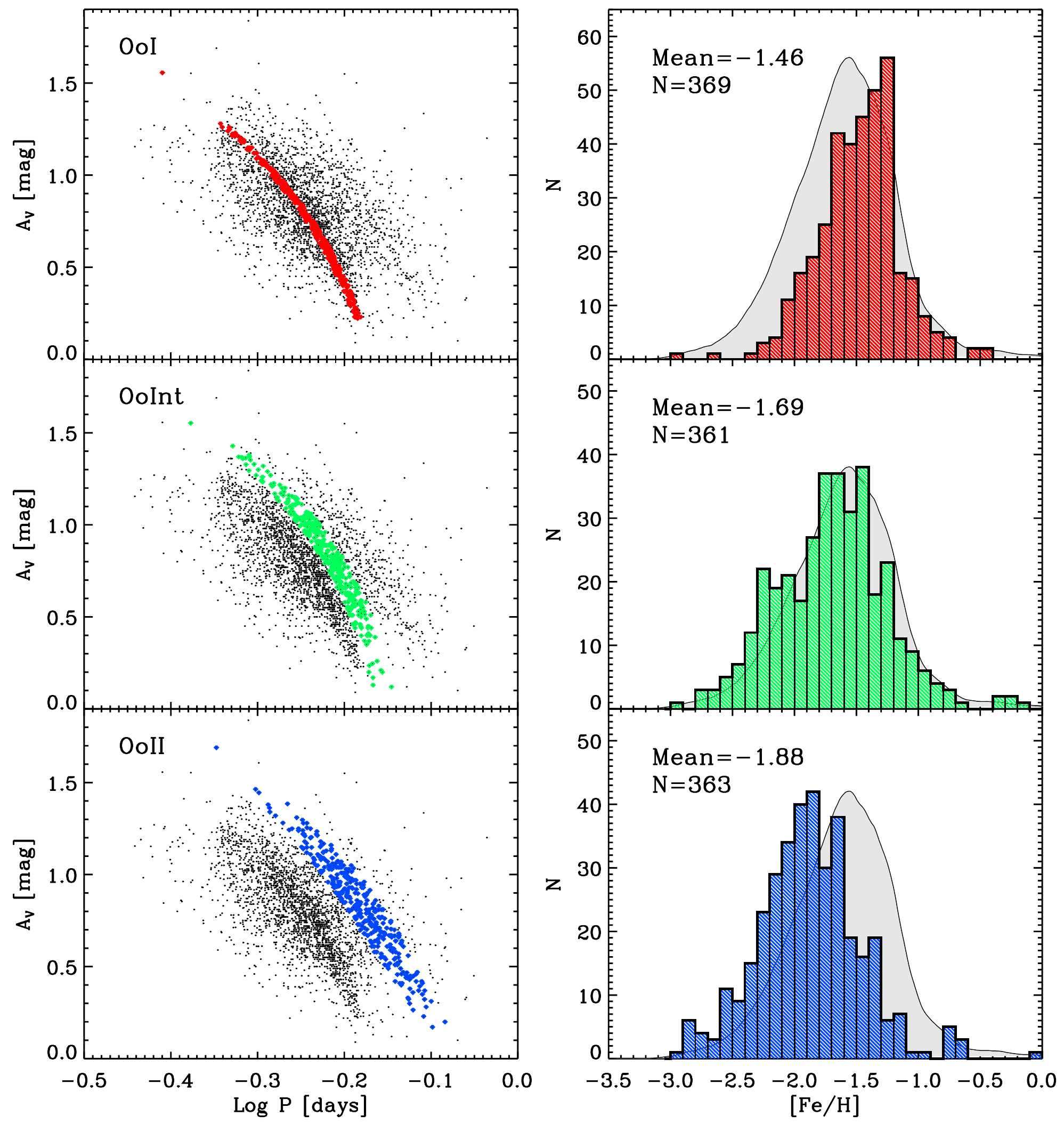} 
\caption{Bailey diagrams (left panels) and metallicity distributions (right panels) for
OoI, OoInt and OoII samples. The grey solid area shows the distribution, 
normalised by the total area, of the 
the entire spectroscopic sample.}\label{fig:ooster}
\end{figure} %%%%%%%

The current iron abundances allow us to investigate the correlation
existing between the Oosterhoff types and metal content. In
Fig.~\ref{fig:ooster} we selected on the Bailey diagram the candidate RRLs 
for OoI, OoII and OoInt around the Oosterhoff loci defined above. 
To overcome possible spurious effects concerning the size of the subsamples, 
the thickness of the regions around Oosterhoff loci were selected in order 
to provide a similar number of RRLs. The left panels of Fig.~\ref{fig:ooster} show the
selections we made for OoI (top in red), OoInt (middle in green)
and OoII (bottom in blue) variables, while the right panels of the same 
figure display the related metallicity distributions over-imposed to the global 
RRLs distribution (grey solid area). The metallicity trend is clear, showing a more
metal-rich distribution for the OoI, with a mean iron abundance of 
\feh=--1.46, to a more metal-poor distribution for the OoII, with a mean iron 
abundance of \feh=--1.88.

The current findings are supporting the empirical evidence concerning the
variation of the mean period as a function of the metallicity brought forward
long ago by \citet{arp55} and by \citet{kinman59}. It is also supporting the period
variation suggested by Sandage in a series of papers
\citep{sandage81,sandage81b,sandage82}. 
However, it is also suggesting that the Oosterhoff dichotomy is caused by the
circumstantial evidence that metal-intermediate Galactic globulars either lack
or only host a few RRLs \citep[the prototype is M13,][]{castellani83,renzini83}.
This means that it is not
directly connected either with an evolutionary or with a pulsation property of
RRLs. The gap in the mean period between OoI and OoII globulars appear to be 
the consequence of the GC diversity.  

It is worth mentioning that the hysteresis mechanism was suggested more than 40
years ago by \citet{vanalbada73} to explain the difference between OoI and OoII
clusters as a variation in the period distribution across the so-called “OR"
region, i.e. the region of the instability strip in which the variables can
pulsate either in the fundamental or in the first overtone or in both of them
\citep{bono94b}. On the basis of the current findings we cannot exclude that the
hysteresis mechanism might affect the period distribution across the instability
strip, but its role in explaining the Oosterhoff dichotomy appears marginal.
These working hypotheses are not new, they were originally suggested by
\citet{castellani83} and \citet{renzini83} in two contributed papers. 

To give them due credits we decided to quote the paragraphs in which they
addressed this specific issue. 

{\em “Concerning the Oosterhoff effect, it now appears that there is a real gap
in \feh\ between the two Oosterhoff types, and the famous discontinuity in
$<$P$_{ab}$$>$ naturally follows from the Sandage's relation: $\Delta \log P
=-0.06 \Delta\feh$. Indeed, BHB clusters with $<$\feh$>$=--1.8, just fill the
gap between Oo. type I and II clusters found by Sandage (cf. Fig.~4 in
\citealt{sandage82}). In other words, the Oosterhoff effect is a consequence of
the non-monotonic behaviour of the HB with respect to \feh."} \citep{renzini83}.

{\em “This does not exclude that an hysteresis mechanism is acting. It only
suggests that at the origin of the different behaviour of the two classes
there is a discontinuity in the evolutionary parameters of the clusters. Either
one accepts a real discontinuity in the history of Galactic GCs, or one
concludes that clusters connecting OoI and OoII do exist, but they have no RR
Lyrae."} \citep{castellani83}.

%_____________________________________________________________________
\section{Stellar population comparisons}

\begin{figure*} %%%%%%%
\centering
\includegraphics[width=1.5\columnwidth]{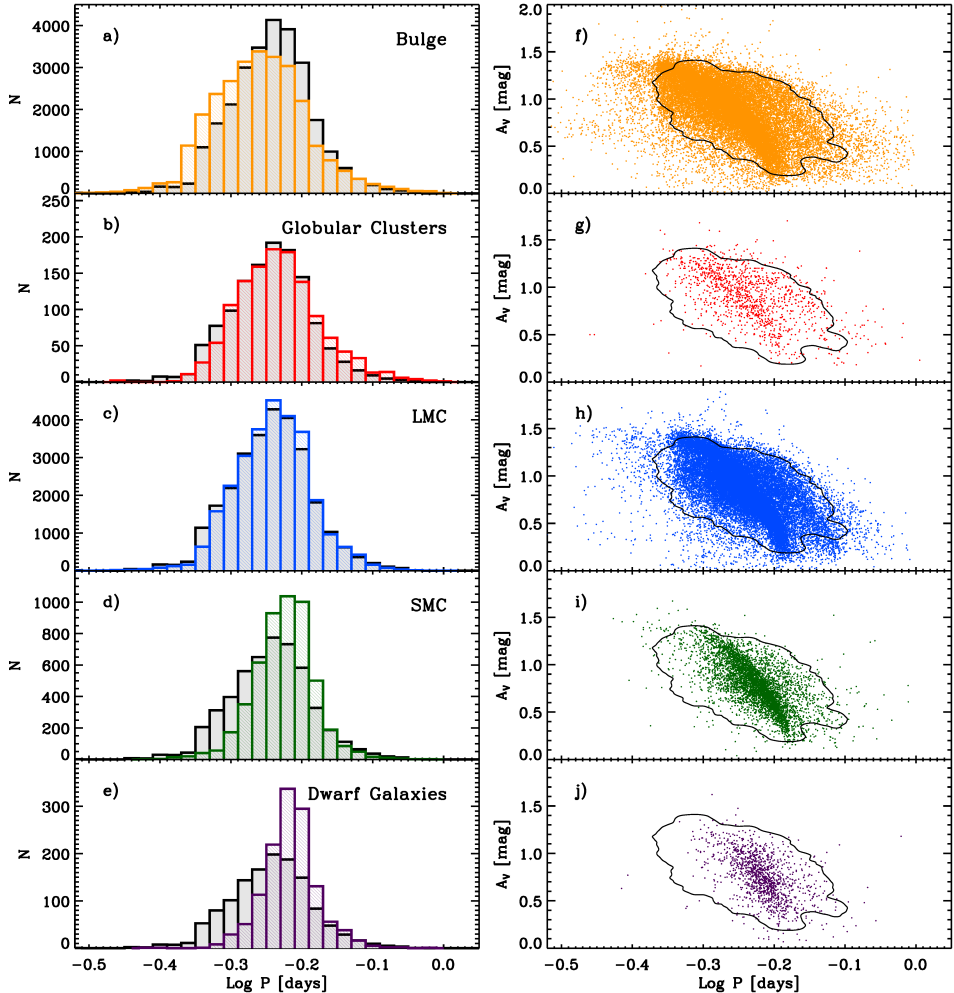} 
\caption{Period distribution (left panels) and Bailey diagrams (right panels) of
the Halo RRab sample compared with different stellar systems. 
The grey contours and histograms show the location and distribution 
(normalised by the total area) of the Halo spectroscopic sample (2,354).}\label{fig:stepop}
\end{figure*} %%%%%%%

The new spectroscopic sample allows us to investigate on a more quantitative
basis the difference in the Bailey diagram and in the period distribution between
Halo RRLs and RRLs in nearby stellar systems. The candidate Halo stars 
(2,354) were separated from the RRLs belonging to the Sagittarius stream 
by using the “spatial over-density" criterium discussed in 
Sect.~\ref{dataset}. Data plotted in the panel a) of
Fig.~\ref{fig:stepop} display the comparison of the period distribution
between the current spectroscopic sample (grey shaded area) and the Bulge RRLs
(OGLEIV, \citealt{soszynski14}, orange shaded area). The panel f) shows the same
comparison, but in the Bailey diagram, and the Bulge RRLs are marked with orange
dots, while the Halo spectroscopic sample with a black contour (95\%\ level).
The empirical evidence indicates that the period distribution of Bulge RRLs is
systematically shorter than Halo RRLs. Moreover, the short period tail is
significantly more relevant in the Bulge sample than in the Halo. This evidence
together with a sizeable sample of HASPs RRLs
\citep{fiorentino15} is suggesting that the metallicity distribution of Bulge
RRLs is systematically more metal-rich than Halo RRLs. 

Panels b) and g) show the comparison between the spectroscopic sample and
cluster RRLs. The difference concerning the occurrence of the Oosterhoff gap has
already been discussed in Sect.~\ref{sec:per-met}. Here we only mention the
large number of metal-rich RRLs present in the field when compared with Galactic
globulars. 
The presence of metal-rich RRLs has been considered for several decades a
“conundrum" \citep{kraft72,taam76, smith84}, because metal-rich (\feh$\ge$--0.7)
globular clusters do not host RRLs. This problem was partially alleviated, by
the discovery of sizeable sample of RRLs in the two metal-rich globulars NGC~6441
and NGC~6338 \citep{pritzl00}. 
The occurrence of RRLs at Solar metal abundance dates back to \citet{smith84}
and to \citet{walker91}, however, these investigations were based on
low-resolution spectroscopy (\deltaS\ method). Only recently, Sneden and
collaborators measured roughly 30 RRLs at Solar metal content by using high
resolution spectra \citep{chadid17,sneden18}. The current findings are soundly
supporting this result and indicate that the lack of RRLs in Bulge, metal-rich
globulars is mainly an observational bias. The next {\it Gaia} data release,
including accurate estimates of both proper motion and geometrical distances,
will allow us to shed new lights on the possible occurrence of metal-rich
cluster RRLs.       

The agreement between the current sample and Large Magellanic Cloud (LMC) RRLs
(panels c) and h)) is quite interesting and fully supports the results obtained
by \citet{fiorentino15,fiorentino17} based on the entire sample of Halo RRLs
known at that time ($\sim$45,000). They found a strong similarity both in the
period distribution and in the Bailey diagram between Halo and LMC RRLs and
suggested that this is a sound independent support for the major merging
scenario \citep{tissera14,zolotov09}. It is worth mentioning that the
metallicity distribution of LMC RRLs has been investigated by
\citet{clementini03}. They found an average metal abundance of \feh$\sim$--1.48
and the metallicity distribution ranges from --2.1 to --0.5~dex. The current
similarity between Halo and LMC RRLs is suggesting that the latter sample might
cover a broader metallicity range.

The comparison between the spectroscopic sample and the Small Magellanic Cloud (SMC)
RRLs (panels d) and i)) shows quite clearly that the former sample includes a
tail of metal-rich RRLs that is not present at all in the SMC. Indeed, the lack
of HASPs is evident both in the period distribution and in the Bailey diagram.
There is only one SMC globular with an age larger than ten Gyrs hosting RRLs,
that is NGC~121 \citep{walker88,fiorentino08} and it is once again metal intermediate
(\feh$\sim$--1.28, \citealt{dalessandro16}).The difference between SMC and LMC
is expected, since the former stellar system is significantly less massive than
the latter one. This means that the chemical enrichment has been less efficient
in the SMC than in the LMC. This is a consequence of the Mass-Metallicity scaling 
relation \citep{chilingarian11}. 

The role played by the total baryonic mass in the chemical evolution becomes
even more relevant in the comparison with RRLs in nearby gas poor dwarf galaxies
(panels e) and j) purple shaded area). 
The RRLs in gas-poor dwarf galaxies adopted in the current investigation 
come from the same sample selected by \citet{braga16}.
These stellar systems only include a handful of HASPS,
i.e. RRab with periods shorter than $\approx$0.5 days. The lack of a sizeable
sample of long-period, metal-poor RRLs is also quite clear. This is the double
circumstantial evidence causing gas poor dwarf galaxies to be “Oosterhoff
intermediate". We have already discussed in Sect.~\ref{sec:intro} the
metallicity distribution of nearby dwarf galaxies, but we would like to add a
few words of caution in using it. Current spectroscopic measurements mainly rely
on high/medium resolution spectroscopy of red giants (APOGEE, \citealt{majewski17}, 
GALAH \citealt{buder18}). The spectroscopic measurements for the stellar 
systems with multiple star formation
episodes are an average of old- and intermediate-age stellar populations. This
is a consequence of the so-called age-metallicity degeneracy along the red-giant 
branch. The consequence is that red-giant stars with old/intermediate-age progenitors 
and different metallicities attain similar magnitudes and colours along the red giant branch. 
A novel approach to overcome this problem
was recently suggested by \citet{monelli14} based on a photometric index
\cubi=[(\uu--\bb)--(\bb--\ii)], but it has only been applied to the Carina
dwarf spheroidal galaxy \citep{fabrizio16} and indicates that the peak of the old stellar
population associated with RRLs is systematically more metal-poor than the
intermediate-age one associated with red clump stars. New and accurate
spectroscopic measurements are required to fully investigate the RRLs in
gas-poor dwarf galaxies. In passing, we also note that nearby dwarf galaxies
always host RRLs and the morphology of the HB is dominated neither by
hot/extreme HB stars nor by red HB stars \citep{bono16}. Moreover, there is no
evidence of the occurrence of a “second parameter" problem among gas-poor dwarf
galaxies.

%_____________________________________________________________________
\section{Conclusions}

The last twenty years have been quite crucial for the understanding of stellar populations 
and evolutionary properties of low-, intermediate- and high- mass stars. 
This relevant step forward applies not only to Galactic stellar populations, but also to 
resolved stellar populations in Local Group and Local Volume galaxies. 
In spite of these indisputable advantages, several long-standing astrophysical problems still 
await for a quantitative explanation of the different physical mechanisms and input parameters 
driving their occurrence. The Oosterhoff dichotomy is among them. Several working hypotheses    
have been suggested in the literature, but the lack of accurate and homogeneous metal 
abundances hampered a solid explanation for the occurrence of this phenomenon. In particular, 
we were lacking firm clues concerning the role that the environment plays in explaining the 
basics of the Bailey diagram. In this investigation we estimated new and homogeneous iron 
abundances for a sample of 2,382 field RR Lyrae stars by using medium resolution 
SDSS-SEGUE spectra. 
They were complemented with estimates available in the literature and based either on 
high, or on intermediate or on low spectral resolutions. We ended up with a sample of 2,903 RRLs, 
the largest and most homogenous sample of iron abundances ever estimated for 
fundamental RRLs. The results we found are summarised in the following.

$\bullet$ The \deltaS\ approach adopted to derive the iron abundances was also validated 
for a fundamental field RRL (V~Ind) for which we collected X-shooter spectra covering the 
entire pulsation cycle. The iron estimates agree, within the errors, on the whole pulsation 
period, including also the critical part of the rising branch. 

$\bullet$ We found a metallicity distribution slightly skewed
toward the metal-poor regime, with a mean iron abundance of \feh=--1.59$\pm$0.01 and a 
dispersion of 0.43~dex.

$\bullet$ The RRL plotted in the Period-Amplitude plane (Bailey diagram) allow us to define the
period-amplitude relations for the three Oosterhoff sequences (OoI, OoII and
OoInt) and to confirm the differences in metal content among these groups.
Indeed, the OoI show an iron distribution more metal-rich (\feh=--1.46) than the
OoInt (--1.69) and the OoII (--1.88). 

$\bullet$ We were able to find a continuous and linear correlation between the metallicity
and the period, confirming the theoretical and empirical evidence brought forward in the literature, 
indicating that the long-standing problem of the Oosterhoff dichotomy among Galactic globulars is 
the consequence of the lack of metal-intermediate clusters hosting RRLs.

$\bullet$ We compared the Halo RRLs period distribution and Bailey diagram with
those of the nearby stellar systems. In particular, the Galactic Bulge and dwarf galaxies differ 
from the Halo, suggesting a metallicity distribution more metal-rich for Bulge stars 
against a more metal-poor distribution for dwarf galaxies.

In this context it is worth mentioning that the analytical relations we are 
providing for OoI, OoII and OoInt groups shall be applied to the mean period 
of sizeable RRL samples. The standard deviation in metal content, at fixed 
pulsation period, is too large to be applied to individual RRLs.
The above findings indicate that the new spectroscopic sample is crucial to 
address a long-standing astrophysical problem. However, they should be cautiously 
treated, indeed, the current analysis is only based on fundamental RRab 
variables. A more comprehensive empirical scenario awaits for spectroscopic 
abundances of first overtone RRc variables.

%_____________________________________________________________________
\vspace{5mm}\acknowledgments
It is a real pleasure to thank the anonymous referee for her/his 
pertinent suggestions that improved the content and the readability of 
the paper.
This work has made use of data from the European Space Agency (ESA) mission {\it
Gaia} (\url{https://www.cosmos.esa.int/gaia}), processed by the {\it Gaia} Data
Processing and Analysis Consortium (DPAC,
\url{https://www.cosmos.esa.int/web/gaia/dpac/consortium}). Funding for the DPAC
has been provided by national institutions, in particular the institutions
participating in the {\it Gaia} Multilateral Agreement.
This research has made use of the GaiaPortal catalogues access tool, ASI - Space
Science Data Center, Rome, Italy (\url{http://gaiaportal.ssdc.asi.it}). We would
also like to acknowledge the financial support of INAF (Istituto Nazionale di
Astrofisica), Osservatorio Astronomico di Roma, ASI (Agenzia Spaziale Italiana)
under contract to INAF: ASI 2014-049-R.0 dedicated to SSDC. 
M.M. was partially supported by the National Science Foundation under Grant No. AST1714534.

\bibliography{biblio}

%___________________________________________________________________

\end{document}